\documentclass[article]{aastex63}
\usepackage{float}
\usepackage{amsmath}
\usepackage{comment}
\usepackage{graphicx}
\usepackage[caption=false]{subfig}
\usepackage[flushleft]{threeparttable}
\usepackage{changepage}
\usepackage{sidecap}
\sidecaptionvpos{figure}{c}

\usepackage{kpfonts}
\begin{document}

\title{Toward Exoplanet Transit Spectroscopy Using JWST/MIRI's Medium Resolution Spectrometer}

\shorttitle{Transit Spectroscopy Using the MRS}
\shortauthors{Deming et al.}

\correspondingauthor{Drake Deming}
\email{ldeming@umd.edu}

\author[0000-0001-5727-4094]{Drake Deming}
\affiliation{Department of Astronomy, University of Maryland,
   College Park, MD 20742, USA}
\affiliation{NASA Astrobiology Institute's Virtual Planetary Laboratory}   

\author[0000-0002-3263-2251]{Guangwei Fu}
\affiliation{Department of Physics and Astronomy, Johns Hopkins University, Baltimore, MD 21218 USA}

\author[0000-0003-4757-2500]{Jeroen Bouwman}
\affiliation{Max Planck Institute for Astronomy, Königstuhl 17, D-69117 Heidelberg, Germany}

\author[0000-0003-0589-5969]{Daniel Dicken}
\affiliation{UK Astronomy Technology Centre, Royal Observatory Edinburgh, Blackford Hill, Edinburgh EH9 3HJ, UK}

\author[0000-0001-9513-1449]{Nestor Espinoza}
\affiliation{Space Telescope Science Institute, 3700 San Martin Drive, Baltimore, MD 21218, USA}

\author[0000-0002-2041-2462]{Alistair Glasse}
\affiliation{UK Astronomy Technology Centre, Royal Observatory Edinburgh, Blackford Hill, Edinburgh EH9 3HJ, UK}

\author[0000-0002-8963-8056]{Thomas Greene}
\affiliation{Space Science and Astrobiology Division, NASA’s Ames Research Center, M.S. 245-6, Moffett Field, 94035, CA, USA}

\author[0000-0002-7612-0469]{Sarah Kendrew}
\affiliation{European Space Agency, Space Telescope Science Institute, 3700 San Martin Drive, Baltimore, MD 21218, USA}

\author[0000-0002-9402-186X]{David Law}
\affiliation{Space Telescope Science Institute, 3700 San Martin Drive, Baltimore, MD 21218, USA}

\author[0000-0002-0746-1980]{Jacob Lustig-Yaeger}
\affiliation{Johns Hopkins University Applied Physics Laboratory, Laurel, MD 20723}
\affiliation{NASA Astrobiology Institute's Virtual Planetary Laboratory}  

\author[0000-0003-4801-0489]{Macarena Garcia Marin}
\affiliation{European Space Agency, Space Telescope Science Institute, 3700 San Martin Drive, Baltimore, MD 21218, USA}

\author[0000-0001-8291-6490]{Everett Schlawin}
\affiliation{Steward Observatory, University of Arizona, 933 N. Cherry Ave., Tucson, AZ 85721, USA}

\begin{abstract}
The Mid-Infrared Instrument (MIRI)'s Medium Resolution Spectrometer (the MRS) on JWST has potentially important advantages for transit and eclipse spectroscopy of exoplanets, including lack of saturation for bright host stars, wavelength span to longward of 20 microns, and JWST's highest spectral resolving power.  We here test the performance of the MRS for time series spectroscopy by observing the secondary eclipse of the bright stellar eclipsing binary R Canis Majoris.  Our observations push the MRS into saturation at the shortest wavelength, more than for any currently known exoplanet system.  We find strong charge migration between pixels that we mitigate using a custom data analysis pipeline.  Our data analysis recovers much of the spatial charge migration by combining detector pixels at the group level, via weighting by the point spread function.  We achieve nearly photon-limited performance in time series data at wavelengths longward of 5.2 microns.  In 2017, Snellen et al. suggested that the MRS could be used to detect carbon dioxide absorption from the atmosphere of the temperate planet orbiting Proxima Centauri.  We infer that the relative spectral response of the MRS versus wavelength is sufficiently stable to make that detection feasible.  As regards the secondary eclipse of this Algol-type binary, we measure the eclipse depth by summing our spectra over the wavelengths in four channels, and also measuring the eclipse depth as observed by TESS.  Those eclipse depths require a temperature for the secondary star that is significantly hotter than previous observations in the optical to near-IR, probably due to irradiation by the primary star.  At full spectral resolution of the MRS, we find atomic hydrogen recombination emission lines in the secondary star, from principal quantum levels $n=$\,7, 8, 10, and 14.

\end{abstract}


\section{Introduction}\label{sec: introduction}

\subsection{Exoplanet Transit Spectroscopy Using JWST}\label{sec: transits_jwst}

Prior to the launch of JWST, exoplanet investigators projected significant scientific results on transiting planets using spectroscopy and photometry from JWST \citep{deming_2009, kendrew_2015, greene_2016, greene_2017, batalha_2018}.  JWST's spectroscopy of transiting planets has indeed been successful at transit \citep{Fu_2022, Fu_2024, ahrer_2023, alderson_2023, dyrek_2024b, esparza-borges_2023, feinstein_2023, lustig-yaeger_2023, madhusudhan_2023, moran_2023, rustamkulov_2023}, at secondary eclipse \citep{august_2023, coulombe_2023, greene_2023}, and in phase curves \citep{bell_2024, kempton_2023, mikal-evans_2023}.  Collectively, this work has used multiple instruments, including NIRISS/SOSS \citep{doyon_2012, albert_2023, Fu_2022, feinstein_2023, holmberg_2023, madhusudhan_2023, radica_2023, taylor_2023}, NIRSpec \citep{jakobsen_2022, birkmann_2022, alderson_2023, august_2023, esparza-borges_2023, espinoza_2023, madhusudhan_2023, moran_2023, rustamkulov_2023}, the NIRCam grisms \citep{greene_2017, ahrer_2023, Fu_2024}, NIRCam photometry \citep{schlawin_2023}, the MIRI Low Resolution Spectrometer \citep{kendrew_2015, bouwman_2023, bell_2024, dyrek_2024b}, and MIRI filter photometry \citep{greene_2023}.  

\subsection{The MIRI Medium Resolution Spectrometer}\label{sec: MRS}
MIRI's Medium Resolution Spectrometer (the MRS, \citealp{Wells_2015, labiano_2021, argyriou_2020a, argyriou_2020b, argyriou_2021, argyriou_2023, law_2023, wright_2023}) is JWST's highest spectral resolution mode, and it also reaches to a longer wavelength than other JWST spectrometers. The MRS is also an integral-field spectrometer with a field of view that varies from $3.2\times3.7$ arcsec at 4.9\,$\mu$m, to $6.7\times7.7$ arcsec at 28\,$\mu$m, and it is spatially undersampled \citep{Wells_2015}. It has been used to obtain spectra of spatially resolved exoplanets \citep{miles_2023}, and projected to characterize planets orbiting white dwarfs using the combined light of the star and planet \citep{limbach_2022}.  The MRS also has important applications in probing protoplanetary disks \citep{arulanantham_2024, sellek_2024, henning_2024}. 

\citet{snellen_2017} described a method to detect carbon dioxide absorption in the nearby (non-transiting) temperate exoplanet Proxima Centauri\,b, using the MRS.  Indeed, the MRS has great potential for spectroscopy of exoplanets in combined light (star+planet), but that potential has yet to be fully realized.  Only one Cycle-1 program (1633), and no Cycle-2 or Cycle-3 programs, use the MRS for transiting planets.  The MRS is a complex instrument that exhibits some pronounced systematic effects such the brighter-fatter effect (BFE) in its detectors \citep{argyriou_2023}. Moreover, the MRS requires using full-frame mode to read the detectors (no subarray modes), and three grating settings in different visits to span the full wavelength range. Those properties may have inhibited proposals to use the MRS for transit spectroscopy.

In this paper, we use the MRS to obtain and analyze time series observations of the classic stellar eclipsing binary system R~Canis Majoris \citep{sawyer_1887}, as a proxy for a transiting exoplanet. Because this stellar binary is bright and has some well determined physical properties, it is a good "ground-truth" test for transit spectroscopy using the MRS.  Our analysis also uses some similar observations of the archetype transiting exoplanet HD\,189733b (program 1633, D. Deming is P.I.).  The goal of this paper is primarily technical and science-enabling, but we also derive new scientific properties of R~CMa. 

\subsection{Organization of this Paper}\label{sec: organization}
This paper is organized as follows.  \S\,\ref{sec: advantages} describes the potential advantages of the MRS for transit spectroscopy.  \S\,\ref{sec: observations} describes our target and the observations. Analyzing the data to extract transit spectroscopy is described in \S\,\ref{sec: data_anal}, starting with an overview of the process (\S\ref{sec: overview}), followed by a group-level analysis to extract spectra (\S\,\ref{sec: anal_groups}), defringing (\S\,\ref{sec: defringe}), derivation of white-light eclipse curves (\S\,\ref{sec: white_light}), solving for the eclipse spectrum (\S\,\ref{sec: solving}), a comparison to the calibration pipeline (\S\,\ref{sec: compare}), and the noise properties of the data (\S\,\ref{sec: noise}).  Our results and their implications are given in \S\,\ref{sec: results}, including implications for exoplanets (\S\,\ref{sec: planet_implications}), and a broadband secondary eclipse and hydrogen emission lines in R~CMa (\S\,\ref{sec: RCMa_implications}).   Our conclusions are summarized in \S\,\ref{sec: summary}.

\section{Advantages of the MRS \label{sec: advantages}}. There are at least three advantages to using the MRS for transit spectroscopy, especially for planets with bright stellar hosts:

\begin{itemize}
\item{ {\bf Reduced sensitivity to saturation.}  Many of the most interesting transiting exoplanets orbit bright stars, wherein JWST's instrumental modes can easily saturate due to the significant photon-collection efficiency of the large JWST primary mirror. The MRS has higher spectral resolving power than other instruments on JWST; the greater dispersion results in lower photon fluxes per detector pixel, and thereby not saturating as easily as lower resolution modes. }

\item{ {\bf Wavelength range.}  The MRS enables spectroscopy to beyond 20\,$\mu$m in wavelength, the longest reach of any spectrometer on JWST.  Spectroscopy at long thermal wavelengths can probe emission by temperate exoplanets via their secondary eclipses.  For planets orbiting bright stars, the MRS has the sensitivity to measure the signatures of cloud absorption at long wavelengths, as calculated by \citet{wakeford_2015}.  Silicate clouds have peak absorption near 10\,$\mu$m, and can be detected using MIRI's Low Resolution Spectrometer (the LRS, \citealp{dyrek_2024b}).  However, other cloud compositions are possible that have significant opacity at wavelengths exceeding the range of the LRS (5 to $\sim$\,12\,$\mu$m, \citealp{kendrew_2015}).  For example, CaTiO$_3$ absorption peaks near 14\,$\mu$m (Figure~9 of \citealp{wakeford_2015}), and that is beyond the range of the LRS, but observable using the MRS. }

\item{  {\bf Potential for cross-correlation detections.} Among the first methods proposed to detect molecules in both transiting and non-transiting close-in planets is the cross-correlation technique.  This method correlates observed high resolution spectra of the star+planet system with a model template of the planet's possible molecular spectrum.  Initial attempts were unsuccessful but promising \citep{wiedeman_2001}, and a successful detection of carbon monoxide by \cite{snellen_2010} led to a flood of ground-based detections, (e.g., \citealp{brogi_2012, birkby_2013} and many others).  A variant of the cross-correlation method has recently been used with transit spectra from JWST/NIRSpec to detect carbon monoxide \citep{esparza-borges_2023, grant_2023}.  The medium spectral resolution of the MRS provides great potential to make similar detections at wavelengths greater than 5\,$\mu$m, as recently demonstrated for directly imaged planets by \citet{patapis_2022}. } 
\end{itemize}

\begin{SCfigure}
\centering
\includegraphics[width=3in]{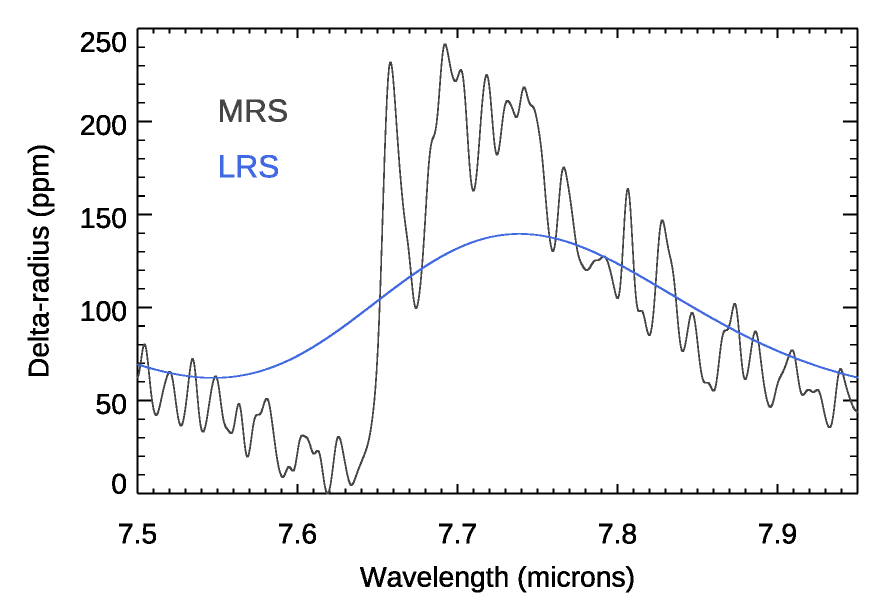}
\caption{Modeled transit spectra of methane in a warm Jupiter at the spectral resolving power of the MRS ($R\sim{3000}$, \citealp{labiano_2021}) and the LRS ($R\sim{100}$, \citealp{kendrew_2015}). The Y-axis is the relative change in transit radius (i.e., depth of the transit) that varies with wavelength. \label{fig: methane} }
\end{SCfigure}

Figure~\ref{fig: methane} illustrates the advantage of the MRS's spectral resolving power, observing the 7.8\,$\mu$m band of methane, calculated for a warm Jupiter's atmosphere in transit.  The resolving power of the MRS varies with wavelength, from $\sim\,3500$ at 5\,$\mu$m to $\sim\,1500$ at 28\,$\mu$m.  We calculated the spectrum in Figure~\ref{fig: methane} at sub-Doppler resolution, and convolved it to the resolving power of the LRS and the MRS.  For the LRS, this band would cover only $\sim 3$ resolution elements.  While that may indeed be detectable, features that occupy only a few data points can be problematic.  The MRS resolves structure in the band, and (for bright systems) could potentially detect this band in either absorption (transit) or emission (eclipse). Except for the very brightest systems, the MRS is unlikely to have a sufficient signal-to-noise ratio to detect this band at full resolution. Spectral rebinning of data from the MRS would allow spectral resolution to be traded for increased sensitivity.  Moreover, cross-correlation detections are likely to be successful using the MRS at full resolving power (especially for eclipses), and would be very complementary to measurements using the LRS.

\section{Observations}\label{sec: observations}

In this work we use observations of R~Canis Majoris during secondary eclipse, and also HD\,189733b during transit. The secondary eclipse of R~CMa is significantly weaker than its primary transit, and provides an appropriate test for exoplanet transits. The observations are summarized in Table~\ref{table: obs}.  R~CMa is bright (V=5.7), and that was an important criterion for us because we aim to obtain an excellent photon-limited signal-to-noise ratio (SNR), and thereby define and investigate any instrumental effects in the MRS that may complicate future transit spectroscopy.  

R~CMa was discovered by \citet{sawyer_1887}, and it has been the subject of numerous studies for well over a century (e.g., \citealp{chandler_1887, guinan_1977, budding_2011, lehmann_2018}, and many other papers).  This Algol-type binary is comprised of an F2III primary, and a G8IV secondary that fills its Roche lobe and probably is transferring mass to the primary star. 

Due to data volume limitations in APT, (the Astronomers Proposal Tool), we observed the secondary eclipse in two separate visits: ingress and egress (Table~\ref{table: obs}).  To obtain a full spectrum using the MRS requires 3 grating settings \citep{Wells_2015}, but we used only the 'A' grating setting ('SHORT' in APT) because that was sufficient for our technical purpose.  There are two other grating settings providing different wavelength channels, but the properties of the wavelength channels that we used are summarized in Table~\ref{table: channels}. We planned the timing of the observations by consulting a historical ephemeris \citep{guinan_1977}, that we updated using low-noise photometry from TESS.

As will be apparent in \S\,\ref{sec: anal_ramp}, we found it informative for technical reasons to also examine data for a system that is not as bright as R~CMa.  We used one transit of HD\,189733b (K2V, V$_{mag}=7.65$), as summarized in Table~\ref{table: obs}. 

\begin{table}
\caption {Summary of the observations.  All of the observations used 30 groups per integration, in the FASTR1 mode.  N$_{integ}$ is the number of integrations.}
\begin{tabular}{llllll}
Event &  Program ID & Date & Start (UTC) & Total Duration (min) & N$_{integ}$  \\
\hline 
\hline
R CMa ingress &  1556 &  2/25/2023  &  13h20m &  291 & 203  \\ 
R CMa egress  &  1556 &  3/15/2023  &  21h44m &  291 & 203 \\
HD\,189733b transit & 1633 & 4/24/2023 & 03h58m & 380 & 257 \\
\hline
\end{tabular}
\label{table: obs}
\end{table}

\begin{table}
\caption {Properties of the four channels that we use in this paper, for grating setting 'A' (SHORT in APT). $N_{sl}$ is the number of spatial slices that we analyze in each channel. The range in wavelength and spectral resolving power are also indicated for each channel.} 
\begin{tabular}{lllll}
Channel &  Band & $N_{sl}$ & Wavelength(s) ($\mu$\,m) & Resolving power \\
\hline 
\hline
1 & 1SHORT & 4 &  4.90-5.74  &  3320-3710 \\ 
2 & 2SHORT & 3 &  7.51-8.77  &  2990-3119 \\
3 & 3SHORT & 2 &  11.55-13.47  &  3320-3710 \\ 
4 & 4SHORT & 1 &  17.70-20.95  &  2990-3119 \\
\hline
\end{tabular} 
\label{table: channels}
\end{table}

\section{Data Analysis}\label{sec: data_anal}

\subsection{Overview}\label{sec: overview}

\begin{figure}
\centering
\includegraphics[width=6in]{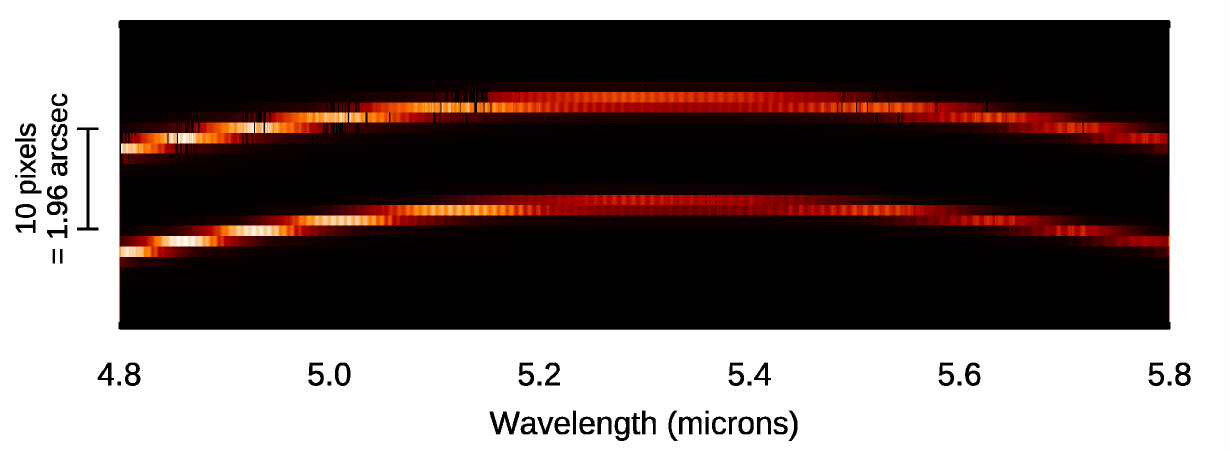}
\caption{\label{fig: compare_data} Illustration of the brightest spectral slice on the detector for channel-1 of R~CMa, using a randomly selected integration in the eclipse ingress visit.  The pixel grid has been resampled and expanded by a factor of 10 in the spatial direction (Y-axis here), to better illustrate the structure of the spectral slice.  The slices are curved on the detector, with the spatial center of each slice varying gradually with wavelength by up to 15 pixels. The upper slice is from the rateints file produced by the calibration pipeline, the lower slice is the same data after our jump corrections are applied using the ramp file (see text).  The slices from the two analyses are displayed together by cutting and joining them into one image, offset in Y by an arbitrary amount, for clarity. The dropouts to zero brightness in the upper slice are NaN pixels in the rateints file.  A prominent effect in brightness is the variation in the illumination of the pixel grid due to different sampling of the PSF as a function of wavelength.  The fringes are also visible as a more subtle pattern of brightness fluctuation versus wavelength. 
}
\end{figure}

We here give an overview of our data analysis process, that uses a combination of the MRS calibration pipeline\footnote{https://jwst-pipeline.readthedocs.io/en/latest} (\citealp{labiano_2016}, and hereafter the pipeline), and our own custom procedures.  Normally, we would begin this process using the 'rateints' files produced by the pipeline (e.g., \citealp{Fu_2024}).  Those files give the rate of electron accumulation in each pixel, after the pipeline solves for the slope of the integration ramp.  However, we found a large number of 'NaN' pixels in the rateints files, where the pipeline was unable to solve for an electron accumulation rate, especially for the most intensely illuminated pixels.  Figure~\ref{fig: compare_data} shows the brightest spectral slice of a rateints file for a randomly selected channel-1 integration in the ingress visit, compared to the same spectral slice from our analysis (as described below).  In addition to the large number of 'NaN' pixels in the rateints file, these data also exhibit a prominent brighter-fatter effect.  We want to correct for the BFE charge migration in a manner that is not possible using the pipeline (explained below).

We therefore begin our analysis with the 'ramp' files produced by version 1.13.3 of the pipeline (using CRDS pmap number 1225).  These files give the integration ramps for each pixel, after several corrections made by the pipeline (starting with the 'uncal' files).  Those corrections include dark current subtraction, nonlinearity correction, and several other steps (MIRI EMI correction, reset anomaly correction, last frame correction, reference pixel correction, and saturation checking). The pipeline also flags jumps produced by cosmic rays, but does not correct them; instead, we correct the cosmic ray jumps in the ramp files (described in \S\ref{sec: anal_ramp}).  

Our analysis solves for electron accumulation rates by analyzing the integration ramps (\S\,\ref{sec: anal_groups}) using the ramp files.  We illustrate the occurrence of jumps, and the BFE, in \S\,\ref{sec: jumps}.  Then, we extract spectra (\S\,\ref{sec: extracting}) by correcting the jumps, and we deal with the BFE by combining pixels at the group level, and then solve for the slope of the combined ramp (\S\,\ref{sec: anal_ramp}).

After extracting spectra, we correct for fringing using a custom procedure that is tuned to these specific data (\S\,\ref{sec: defringe}).  We make comparisons to a conventional analysis using the pipeline with respect to linearity and defringing in \S\,\ref{sec: linearity} and \S\,\ref{sec: compare_defringe}, respectively.

\subsection{Analysis of the Integration Ramps}\label{sec: anal_groups}

\subsubsection{Jumps and the Brighter-Fatter Effect}\label{sec: jumps}

\begin{figure}
\centering
\includegraphics[width=6in]{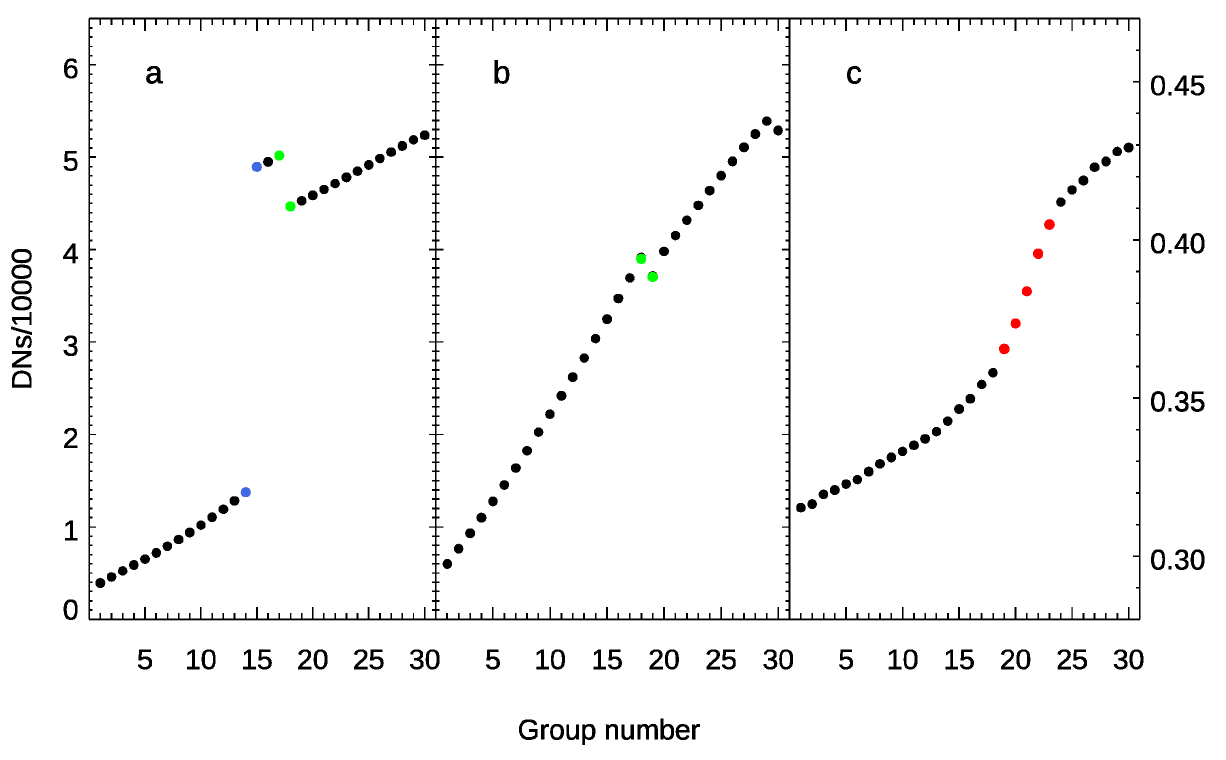}
\caption{Examples of non-ideal ramps in the R~CMa data, using the brightest portion of the spectrum (channel 1 in grating position A/SHORT).  The points in blue, green, and red are real data points, but are plotted in color to call attention to the effects we discuss.  Panels 'a' and 'b' show jumps in the data.  The blue points mark a positive jump due to a cosmic ray hit.  The green points in panels a and b show negative jumps. Panel c shows the brighter-fatter effect; the red points mark an increase in slope due to charge migration from an adjoining pixel.  The Y-axes are in data numbers (DNs) divided by 10,000; read the left axis for panels a and b, panel c axis on the right.}
\label{fig: first_group} 
\end{figure}

MIRI detectors are sampled 'up the ramp', i.e. read non-destructively as electrons accumulate in response to the photon flux on each pixel \citep{ressler_2015, morrison_2023}.  We inspected the signal from individual pixels in the 'ramp' files as a function of group number.  Small samples of randomly selected pixels revealed some non-ideal behavior for the spectrum with the highest flux level (R~CMa, channel 1), illustrated in Figure~\ref{fig: first_group}.  

Figures~\ref{fig: first_group}a and \ref{fig: first_group}b show jumps in the data for representative pixels.  Figure~\ref{fig: first_group}a shows a positive (upward) jump (discontinuity, at the blue points) caused by a cosmic ray hit, that is followed by a negative (downward) jump (green points).  Four points can be made concerning the jumps:
\begin{itemize}

\item The jumps occur in the ramp files without processing of any kind on our part. We find them, but we do not create them. 

\item Using the term jump, as conventionally used for cosmic ray effects, is convenient, but we do not mean that the negative drops in signal are a manifestation of cosmic rays.  Some detector effects such as persistence, and row and column effects, can be negative as well as positive \citep{dicken_2022, dicken_2024}, but their temporal behavior is not purely discontinuous as are these negative jumps.  

\item The negative jumps outnumber the positive jumps by a factor of approximately 8, and they do not occur near the same pixel locations.  Hence the majority of the negative jumps are not associated with cosmic rays.  

\item The negative jumps are {\it not} present in the uncal files that the pipeline uses to produce the ramp data. Hence the negative jumps are created by software, they do not arise in the detector.  The positive jumps do occur in the uncal files, so they are cosmic ray hits that we correct using the ramp files (\S\ref{sec: anal_ramp}). 

\end{itemize}

Below, we have explored the statistics of occurrence for both positive and negative jumps, as an aid to understanding their properties.

In Figure~\ref{fig: first_group}a, a negative jump occurs at a large DN, at approximately 76\% of the hard saturation limit of 65535 DNs, \citep{argyriou_2023}.  The negative jumps tend to occur at high charge levels (see below), but not always.  Figure~\ref{fig: first_group}b shows an example of a negative jump that occurs at a not-yet-saturated charge level ($4\times10^{4}$ DNs).  

Figure~\ref{fig: first_group}c illustrates the brighter-fatter effect, as described for MIRI by \citet{argyriou_2023}. BFE is caused by large differences in the effective bias voltage between adjacent pixels, that cause photoelectrons to migrate.  The amount of charge migration depends on the effective voltage difference between pixels, that in turn depends on the difference in the illumination of adjacent pixels.  That difference is especially great at the shortest wavelengths due to the undersampling of the spatial PSF by the MRS. The large contrast between adjacent pixels in these R~CMa spectra enable a very strong BFE. The BFE illustrated in Figure~\ref{fig: first_group}c is a prominent effect compared to more subtle BFE examples seen in data wherein the PSF is more densely sampled. For example, compare to Figure~5 of \citealp{argyriou_2023} that shows the BFE for imaging data.  The red points in Figure~\ref{fig: first_group}c illustrate the most obvious range of groups where the pixel is receiving charge that migrates from an adjoining pixel, and consequently the slope of the ramp increases.  That increase in slope is temporary: when the adjacent pixel saturates the charge migration apparently stops and the slope returns to a normal value.  It is apparent from Figure~\ref{fig: first_group}c that fitting a linear slope, modified by a saturation term(s), would probably not be successful when the BFE is significant.  Instead, we have dealt with the BFE by combining spatial pixels at the group level, as described in \S\,\ref{sec: extracting}.

\subsubsection{Statistics of the Jumps}\label{sec: stat_jumps}

\begin{figure}
\plottwo{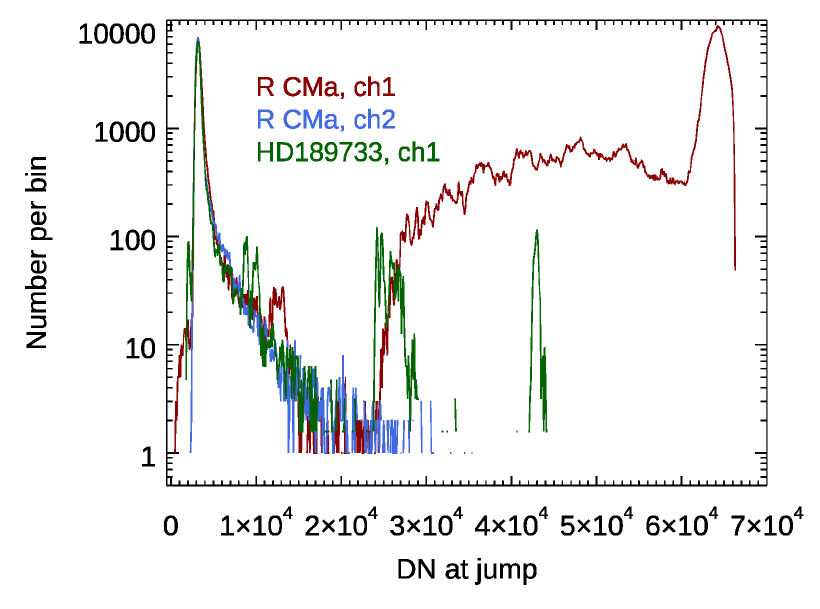}{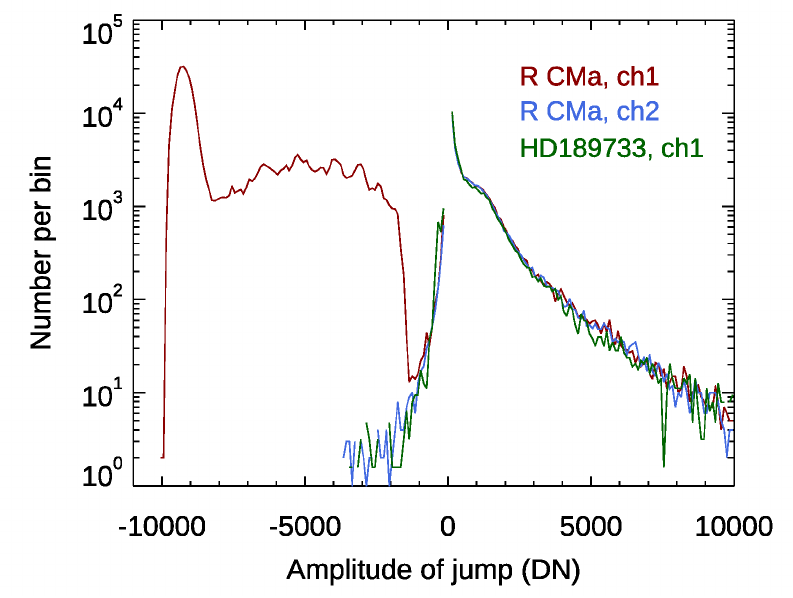}
\caption{Left: histogram of jump occurrence as a function of flux level in DNs for R~CMa channels 1 and 2, and HD\,189733b channel 1.  Right: histograms of the occurrence of jumps as a function of jump amplitude, both positive and negative.  For both panels, the histogram bins have width of 100 DNs.  The total number of pixels in the histograms is $7.3 \times 10^{6}$ for each of the R~CMa spectra, and $4.6 \times 10^{6}$ pixels for the spectrum of HD\,189733. \label{fig: jump_hist} }
\end{figure}

Our analysis detects and corrects both the positive and negative jumps in the ramps (defined and described in \S\,\ref{sec: anal_ramp}), but we want to understand the statistical properties of the jumps, especially the negative jumps. We compiled statistics for the brightest portions of the spectra where the negative jumps are the most numerous: both channels 1 and 2 for R~CMa, and also channel 1 in the HD\,189733b data, shown in Figure~\ref{fig: jump_hist}.  We made histograms of the number of jumps as a function of the flux level (in DN) where the jump occurs, and the number as a function of the amplitude of the jump (both positive and negative jumps).  The IFU produces several spatial slices across the stellar PSF. The range of pixels for the histograms spanned 20 columns centered on the brightest spatial slice of each spectrum, and including detector rows between 50 and 950 (conservatively avoiding rows near the edges of the 1032x1024 detectors), and using all integrations obtained for each target.  That results in $7.3 \times 10^{6}$ pixels for each spectrum of R~CMa (channels 1 and 2), and $4.6 \times 10^{6}$ pixels for channel 1 of HD\,189733.  

The histograms of jumps are shown in Figure~\ref{fig: jump_hist}: the left panel shows the distributions of jumps versus signal level (i.e., DNs), and the right panel shows the number of jumps versus the jump amplitude.  The left panel has two main peaks; a peak at small signal levels occurs for all three spectra because the collection of pixels is dominated by the columns at low signal level adjacent to the brightest spatial slice.  At high signal levels (that occur primarily in channel 1), R~CMa shows a broad peak above 30,000 DNs, with an additional sharp peak at the highest signal levels near 60,000 DNs.  R~CMa dominates the jumps above 30,000 DNs, but HD\,189733b shows a peak in jumps near 45,000 DNs.  The right panel of Figure~\ref{fig: jump_hist} shows the number of jumps as a function of jump amplitude for all three spectra.  Channel 1 of R~CMa dominates the jumps at negative amplitudes.  The other two spectra have primarily positive jumps (consistent with cosmic rays), with only the tail of their distributions extending to negative amplitudes. 

\begin{figure}
\plottwo{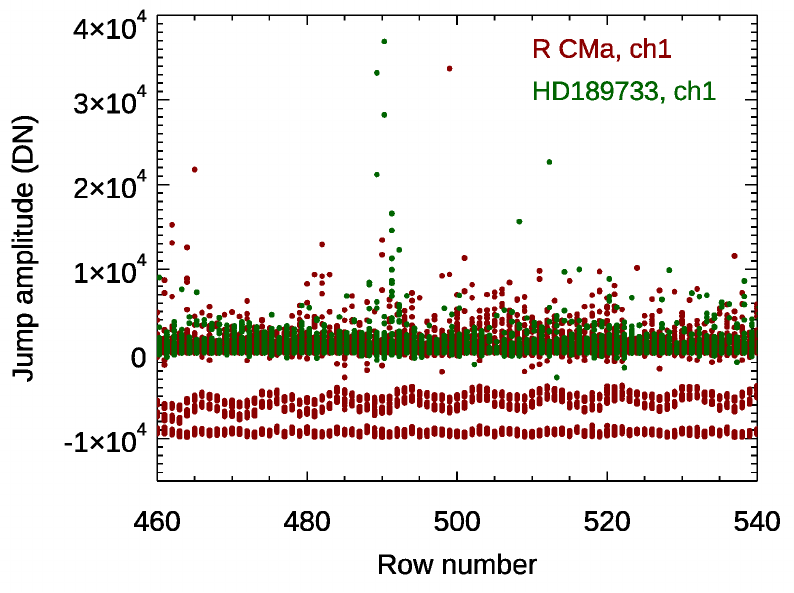}{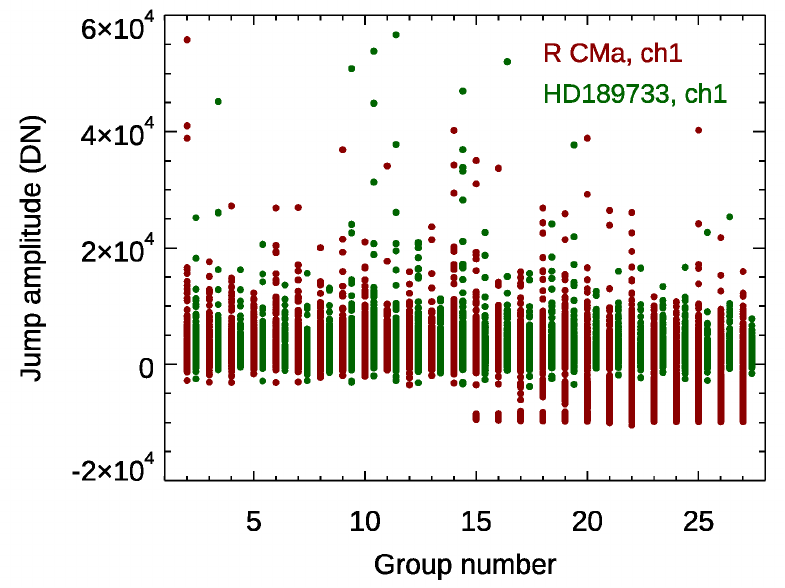}
\caption{Left: Jump amplitude versus row number for the two channel 1 spectra, expanding an interesting portion near the center of the detector.  Each point is a single jump, and many points overlap.  Right: Jump amplitude versus group number for both channel 1 spectra.  Both spectra used 30 groups, but we analyzed only up to group 27, avoiding any last-group effects.  Each point is a single jump, many points overlap but HD\,189733 is offset by +0.4 in group number to minimize overlap. \label{fig: row_group} }
\end{figure}

We also investigated the occurrence of the jumps as a function of row number and of group number, shown in Figure~\ref{fig: row_group}.  We expanded the X-axis for the plot versus row number to illustrate that HD\,189733 has a concentration of positive jumps near row 492. These jumps are concentrated in time (not illustrated) as well as concentrated spatially, and they are potentially due to a cosmic ray shower\footnote{https://jwst-docs.stsci.edu/known-issues-with-jwst-data/shower-and-snowball-artifacts}. A different type of spatial distribution is illustrated by the negative jumps versus row number for R~CMa.  Those jumps show a consistent band near -10,000 DNs, and a quasi-sinusoidal variation in jump amplitude versus row number for jumps with amplitudes of minus a few thousand DNs.  This pattern of negative jumps may be related to the optical fringing that could potentially exacerbate software effects that are related to the source brightness. The right panel of Figure~\ref{fig: row_group} shows that the negative jumps only occur for group numbers exceeding 14, consistent with occurring at large accumulated charge levels.  

The behavior described above, combined with their absence in the uncal files, suggests that the negative jumps are created by a software effect in the pipeline that is related to the signal level or the group number. After further investigation (by D.L.), we conclude that the negative jumps occur because the linearity correction step isn't applied to pixels that are already flagged as saturated. Once a pixel trips the saturation flag in a given group, all groups after that are also flagged as saturated.  Linearity correction can then increase the counts of earlier groups, but won't adjust the saturated groups, leaving a negative jump.  Since saturation in a given pixel is expanded to neighboring pixels to help mitigate BFE, this effect can also occur at lower DN in pixels adjacent to those that reach 'true' saturation.

Analyses that solve for the slopes of integration ramps in individual pixels need not be affected by the negative jumps, because those analyses would normally not use the pixels flagged as saturated.  However, in our analysis we combine pixels at the group level in order to recapture the migrated charge (\S\,\ref{sec: extracting}).  In that process, we could lose information if the ramps for all combined pixels were restricted to groups that were unsaturated for the brightest pixel.  Therefore, we correct the negative jumps in a manner analogous to the positive jumps (\S\,\ref{sec: anal_ramp}), and we deal with saturation in the combined ramp as described in \S\,\ref{sec: saturation}.
   
% In that respect the negative jumps may be similar to the row and column effects described by \citet{dicken_2022}.

Channel 1 of R~CMa observed using 30 groups is a brighter spectrum than needed for any exoplanet host, so the negative jumps should not be a direct problem for exoplanet transits.  Nevertheless, there are other effects such as the BFE that can be expected to be significant in exoplanet transits.  We therefore explore to what degree a low noise eclipse spectrum can be derived for R~CMa in all of these data, including channel 1.  We defer the transit spectrum of HD\,189733b to a future paper focused on that planet. 

\subsubsection{Extracting Spectra}\label{sec: extracting}

We now turn to extracting spectra from the data frames, starting at the group level. The presence of a strong BFE effect in these data (Figure~\ref{fig: first_group}) motivates us to combine pixels at the group level prior to solving for the rate of electron accumulation that indicates the source flux.  If we combine pixels, we can potentially re-capture the charge that migrates in the BFE.  Our procedure combines pixels spatially (columns on the detector), but not combining in the wavelength dimension (rows on the detector).  BFE in the wavelength dimension degrades the spectral resolution (see Figure~8 of \citealp{argyriou_2023}), but our analysis ignores that degradation (it could be modeled when retrieving planet properties from the spectra).

Combining spatial pixels spatially at the group level is straight forward. However, simple addition of pixels is not optimal because it does not sufficiently discriminate against background emission \citep{rigby_2023}, that is significant even for bright sources like R~CMa.  We therefore use the optimal extraction method described by \citet{horne_1986}.  This method weights and integrates the pixels using the spatial point spread function (PSF), thereby maximizing the signal relative to the background noise. This weighting and integration over the spatial PSF may not capture migrated charge as thoroughly as would a straight summation. Nevertheless, it accounts for variation of the PSF with wavelength, and we find that it adequately recovers the charge migration that occurs via the BFE.  This group-level optimal extraction thereby allows us to derive spectra whose noise level is close to the theoretical limit, as we now describe.

For each channel, we begin our analysis with spectral images in group-level data (the 'ramp' files). We subtract the spectral image using the first group from spectral images at all subsequent groups, and we divide the differences by a pixel-flat file (a CRDS reference file).  There are multiple spatial slices that sample the source PSF.  We use the brightest slices that are above the background noise level, ranging from 4 slices in channel-1, to 1 slice in channel-4 (see Table~\ref{table: channels}).  For each spatial slice at each wavelength (varying approximately as row number on the detector) we extract the spatial profile using group 20, and we fit Gaussians to those spatial profiles.  A median filter produces a smooth variation of the spatial center and width of the Gaussian fits as a function of row number, and we use those median values in our optimal weighting.  We experimented with parameters of the median filter and other methods of smoothing the spatial profiles, but we found that the SNR of our extracted spectra did not depend strongly on those parameters.

The wavelengths corresponding to each slice from the IFU are very similar, but with small differences \citep{argyriou_2021}.  We use the wavelength calibration from the 'calints' files, choosing the wavelength values that follow the curvature of each spectral slice. We adopt the wavelength corresponding to the peak brightness of the slice at each row of the detector.  Lines of constant wavelength are not exactly parallel to rows of the detector. Hence, the wavelengths vary across a slice at each row, but that variation is no more than 20\% of the spectral resolution, and we ignore it.  We interpolate the extracted spectra of each slice to the wavelengths of the brightest slice, and (eventually) co-add the spectra.

\subsubsection{Electron Accumulation Rate}\label{sec: anal_ramp}

With the center and width of the Gaussian spatial profile determined, we weight the data number per pixel for each group and extract total electron counts for each group at each wavelength of each spatial slice.  The combination of those electron counts for a given wavelength of a given spatial slice define a ramp function that is analogous to the ramps produced for individual pixels in conventional analysis pipelines.  Like conventional ramps, our ramps are affected by cosmic-ray hits (positive jumps), and also by the negative jumps (\S\,\ref{sec: jumps}). All of those jumps must be corrected prior to fitting to the ramps. We experimented with different correction methods, and our best results are obtained as follows.  For a given ramp, we calculate the group-to-group differences in the electron count and we construct a 5-group running median of those differences.  Our electron counts use a detector gain of 4 electrons per data number\footnote{https://jwst-docs.stsci.edu/jwst-mid-infrared-instrument/miri-instrumentation/miri-detector-overview/miri-detector-performance}.  In practice the gain varies with wavelength, but for simplicity we have adopted a constant value.  A jump is defined by a deviation from the running-median difference whose absolute value exceeds a given threshold, and we set those deviated differences to the running median. A threshold equal to twice the average group-to-group difference, with a minimum of 100 electrons, produced our best results, as judged by the noise level in the extracted spectra. With the sequence of group-to-group differences corrected in that manner, we sum the corrected differences to reconstruct a corrected ramp.

Source fluxes in the extracted spectra are related to the slope of the ramp at each wavelength.  An ideal ramp would be strictly linear.  However, the real detectors have non-linearity and saturation, and other effects such as the BFE \citep{argyriou_2023} and reset charge decay \citep{morrison_2023}.  We tried multiple ramp-fitting methods in order to optimize the signal-to-noise ratio, presuming that all of those effects could remain in the data to some degree, even after correction.  Our methods varied the range of groups to include in the fit, and the order of the fitted polynomial.

\begin{figure}
\centering
\includegraphics[width=3in]{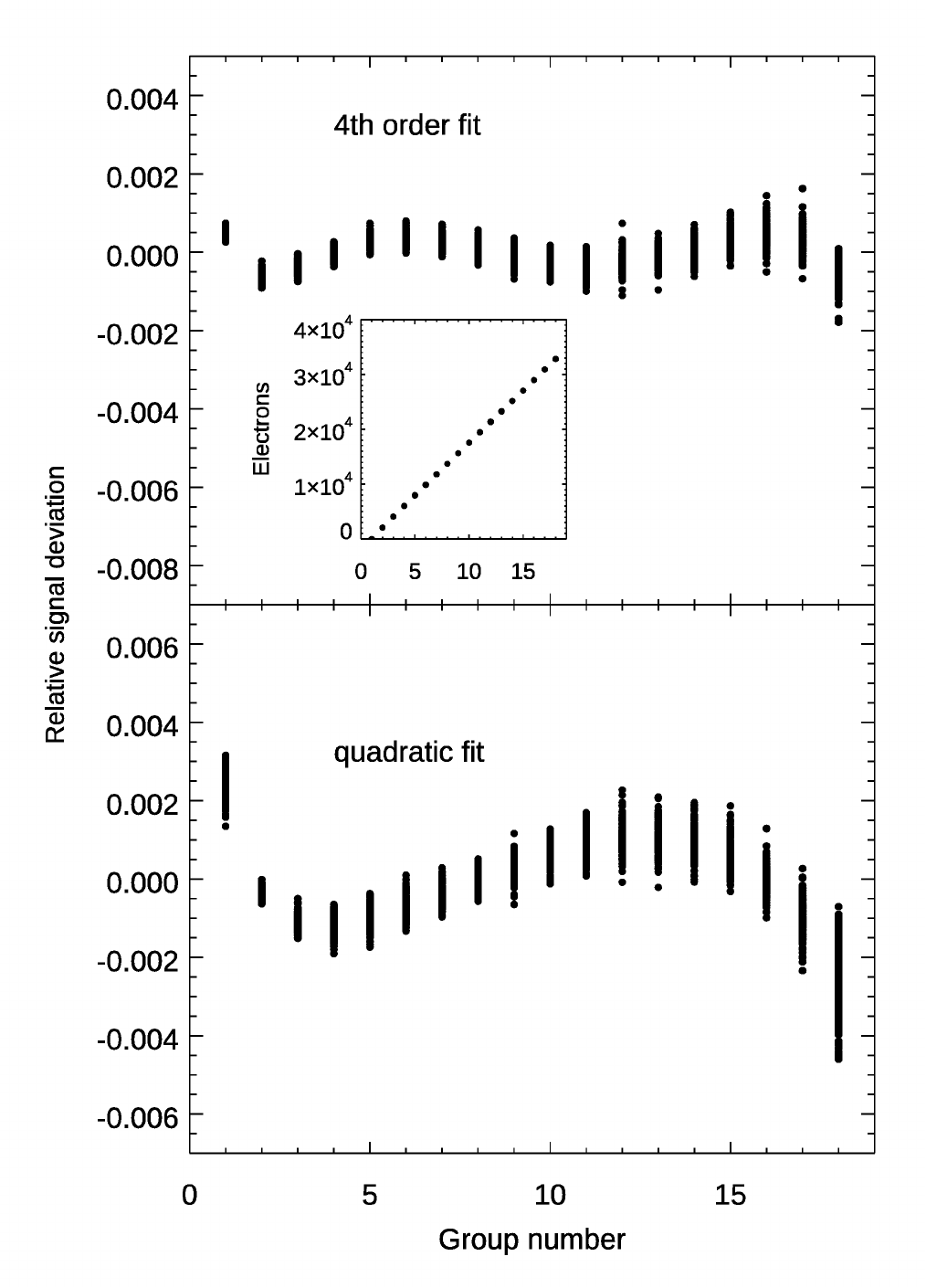}
\caption{\label{fig: ramp_fit} Results of fitting to detector ramps that are integrated over the spatial PSF using optimal weighting, illustrated for the eclipse ingress portion of the channel-1 data.  The inset plot shows the ramp averaged over all wavelengths in channel-1, and also averaged over all integrations in the time series for the eclipse ingress.  The main panels show the fractional differences from fitting the ramp with a quadratic (lower panel) and 4th-order polynomial (upper panel).  Each point in those difference plots is from a single integration in the time series, but averaged over wavelength.}
\end{figure}

We evaluated the results of fitting to the ramps based on the noise level in the extracted spectral flux at several wavelengths in the time series data.  On that basis, omitting the first group is always best, and we also confirmed that the last group is consistently anomalous \citep{morrison_2023}.  Our best results for channel-1 were obtained by restricting the range of groups from 2 to 19 (versus 1 to 30 that were observed).  In channels 2, 3, and 4 we have the best results using groups 2 to 29, and we attribute the need for a more restricted range in channel-1 to saturation (discussed below).  The method of fitting to the ramp is equally important as the range of groups; we tried polynomial fits with degree as high as 4.  Indeed, our best results were obtained with either a 3rd or 4th degree polynomial, that were equivalent to within the round-off precision for all of our results.  For consistency with our fringe removal procedure (\S\ref{sec: defringe}), we used the 4th degree polynomial, as illustrated in Figure~\ref{fig: ramp_fit}.  The small inset portion of Figure~\ref{fig: ramp_fit} illustrates the ramp, averaged over wavelengths and integrations for the channel-1 eclipse ingress data.  While the ramp appears by eye to be quite linear, fitting polynomials (main panels of Figure~\ref{fig: ramp_fit}) shows that a 4th degree polynomial is needed to capture small amounts of curvature in the ramp.

We tried using the linear term in the 4th order polynomial to represent the source rate of electron production, but an alternate formulation gave much better results.  After fitting the polynomial, we subtract the fitted value at the minimum group number from the fitted value at the maximum group number.  That difference (in electrons) is proportional to the electron flux from the source at a given wavelength; it is conceptually similar to simply subtracting the electron number at the smallest group from the accumulated electrons at the largest group.  However, subtracting the fitted values rather than the measured numbers avoids any residual outliers that escape the median-filtering process, and the polynomial fit uses all of the intermediate values of accumulated electrons.

Our use of a 4th degree polynomial could be questioned based on the obvious principle that fitting any data by increasing the number of fit parameters will always produce better fits, without necessarily being justified physically. However, we judge the results not by the quality of the ramp fits {\it per se}, but by the scatter in the time series of the extracted flux levels.  The difference between our best method and more conventional methods (e.g., fitting a quadratic and using the linear term) is not subtle: the scatter in the time series values improves by almost a factor of two.  We conclude that exoplanet transit observers who use the MRS may be able to significantly reduce their uncertainties through analysis of group-level data. 

In parallel with extracting fluxes from the spatial slices of R~CMa, we also extract the background emission (for channels-2, -3, and -4, but not channel-1).  That background includes zodiacal light, and thermal emission at the longest wavelengths.  We extract the background from a region of the detector that views the sky adjacent to R~CMa, but where the flux from R~CMa is not detectable (i.e., no spatially-peaked source).  Adopting the same spatial profile width as for the stellar extractions, we perform an optimal weighting process identical to the stellar extractions to yield a background value at each wavelength, for each integration.  We do not expect - or observe - the background to change during the time series, nor do we expect it to change rapidly with wavelength.  To improve the precision of the background subtraction, we smooth the background over wavelength using a 31-point median, and we average the background spectra over all integrations in each time series (averages for both ingress and egress).  Average background spectra for both ingress and egress are subtracted from each R~CMa spectrum prior to defringing, that we discuss in \S\,\ref{sec: defringe}.

\subsubsection{Saturation}\label{sec: saturation}

The ramp data we use have been corrected for non-linearity by the pipeline, but saturation can still occur.  We did not make an explicit evaluation of detector saturation during the ramp fitting process.  Although that may seem ironic, given the large electron flux produced in our observations, our strategy is consistent with our ramp analysis.  Using our total span method, we found an increase in noise for the extracted spectra at wavelengths shortward of 5.2\,$\mu$m.  That increase in noise is our first indicator that the spectra at the shortest wavelengths are approaching saturation. We checked that by examining the integration ramps as a function of wavelength.  Our ramps comprise a weighted sum of multiple spatial pixels at each wavelength, and weighted sums are not as informative as individual pixels.  Hence we examined the ramps of the brightest pixel in the brightest slice at each wavelength.  In the case of the channel-1 data, those brightest pixels reach the CRDS-defined saturation level of 55500 DNs, for wavelengths less than 5.24\,$\mu$m.  The very brightest pixels reach that saturation level by group 16, but wavelengths longward of 5.24\,$\mu$m do not saturate even at the last group (30). Those properties are in good agreement with the increase in noise, with the statistics of the negative jumps (\S\,\ref{sec: stat_jumps}), and with the range of groups used in our weighted sum ramps (below, and also \S\,\ref{sec: anal_ramp}).  

Given the above discussion, we use a simple threefold strategy to avoid saturation, and optimize the signal-to-noise ratio in the extracted spectra: 1) limit the range of groups used in the ramp solution (e.g., groups 2 to 19 for channel-1, based on the optimal signal-to-noise ratio as per \S\,\ref{sec: anal_ramp}; that implicitly avoids much of the saturation.  2) omit the remaining saturated wavelengths $\lambda\,<\,$5.2\,$\mu$m, and 3) use the total span of the detector ramp at the longer (unsaturated) wavelengths.  Our strategy should be effective for exoplanet transits, because none of those observations will have to acquire as many electrons as in the brightest portions of our R~CMa spectrum.

\subsection{Defringing}\label{sec: defringe}

Layered structures in silicon detectors very commonly cause interference fringes in measured spectra in the infrared \citep{kester_2003}.  The MIRI detectors have fringes \citep{argyriou_2020b, argyriou_2021}, that are spectrally resolved by the MRS, and their properties can vary with how the pixels sample the point spread function (\citealp{gasman_2023}, the MRS is undersampled).  Although the MRS calibration pipeline has a defringing algorithm, we decided to develop a defringing method that is optimized for time series observations such as transits. The referee of this paper points out that there are additional effects beyond optical interference that can contribute to the fringe phenomenon (and we agree). For example, errors in the slopes of the integration ramps may vary with the source flux and thus be coupled to the optical interference, and to in-transit versus out-of-transit flux from the star. We therefore use the term fringes to refer to the totality of those effects, bearing in mind that the underlying optical interference is likely to be intrinsically stable.  If the total fringes were completely stable, it would not be necessary to remove them for a transit, because the fringe response would cancel in the in-transit versus out-of-transit ratio.  However, given the above discussion, it seems prudent to assume that the fringes are not completely stable, so we developed a defringing method that is appropriate for transit spectroscopy, as we now describe.

\begin{figure*}
\centering
\includegraphics[width=3in]{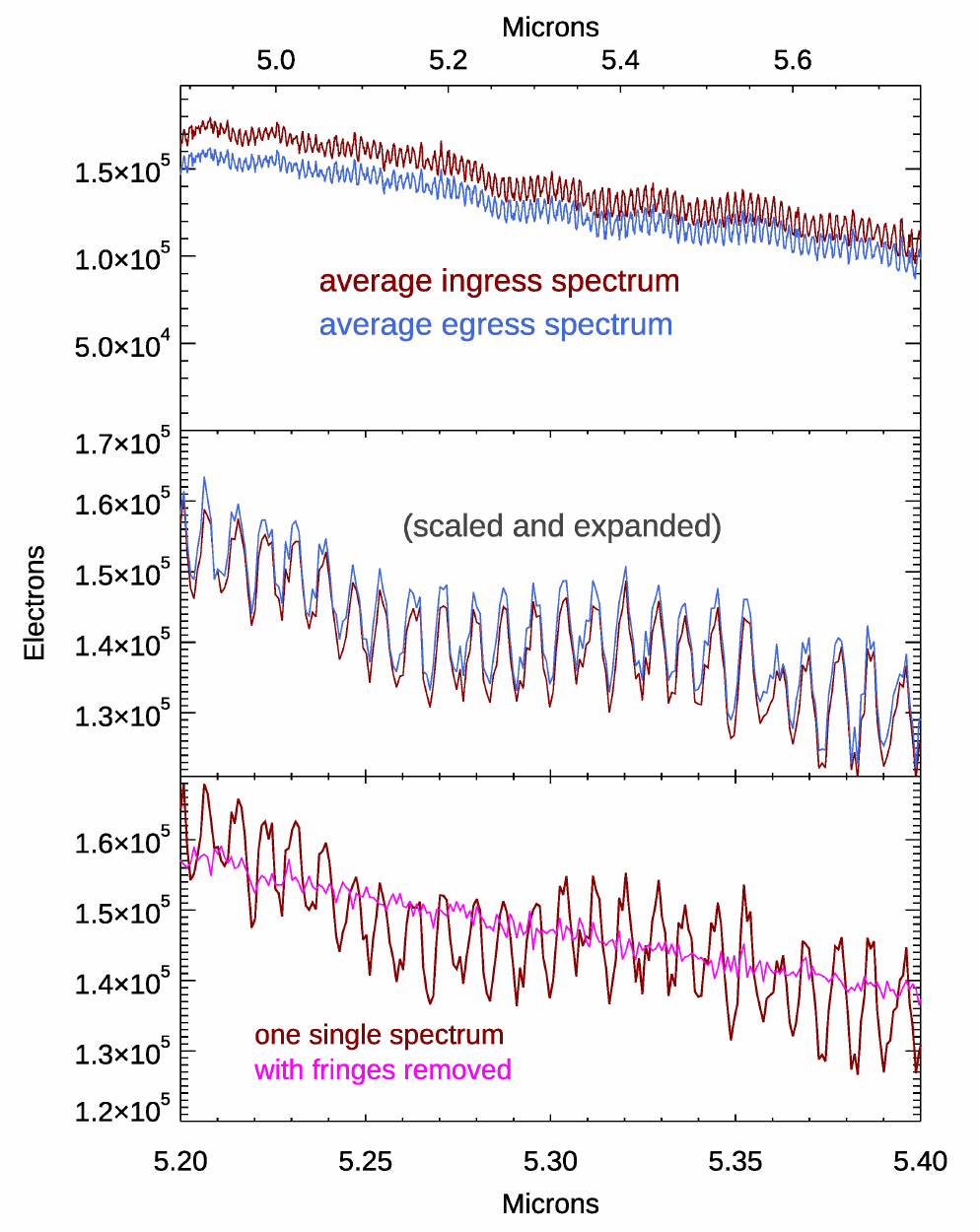}
\caption{Spectral fringes in the MRS spectra of R~CMa.  {\it Top panel (wavelength scale at top):} Average of all integrations for a single spatial slice at both ingress and egress. Note the similarity of the fringe pattern for those two visits, that are separated by 18 days. {\it Middle panel (wavelength scale at bottom):} Expansion of the fringes from the top panel in both wavelength and electron number, and normalization to equal signals, so that the similarity of the fringes at both visits is more apparent.  {\it Bottom panel (wavelength scale at bottom):} Spectrum of a randomly selected single integration during ingress, with the defringed spectrum for that integration overplotted.}
\label{fig: fringes}
\end{figure*}

Our defringing process is illustrated in Figure~\ref{fig: fringes}. The top panel of that figure shows the fringes for a single spatial slice of R~CMa's channel-1, at both ingress and egress, each averaged over all integrations in those visits. The difference in the total flux levels (most evident at the shortest wavelengths) is astrophysical, due to the eclipse.  The fringe structure is very similar for both ingress and egress, in spite of the 18-day difference in the time of the observations.  The middle panel of Figure~\ref{fig: fringes} expands the central portion in wavelength, and normalizes the spectra to the same total flux, to better illustrate the similarity of the fringe pattern. We find (in $\sim$\,agreement with previous work) that there are fringes whose periodicity is both short ($\sim$\,10 pixels) and long ($\sim$\,200 pixels). The long duration fringes are believed to be due to the reflectivity of the buried contact \citep{argyriou_2020a}.  Nevertheless, we treat both the short- and long-period periodic structures in the spectra as instrumental effects to be removed.  Also, as noted above, we emphasize that we use the term 'fringes' very broadly to include the totality of instrumental and data effects that create quasi-periodic fluctuations in the extracted source flux versus wavelength.

Our defringing method is based on the principle (often invoked in transit analyses) of using the data itself to calibrate the instrumental effect (i.e., remove the fringes).  One simple method would be to divide the spectra at each integration by the average of all spectra for that visit.  However, that assumes that the fringes are stable, and we want to avoid that assumption.  Instead we use a higher-order version of that principle.  We first remove the continuum from each spectrum (i.e., single integrations), by fitting a polynomial to the flux versus wavelength, and dividing by it, leaving the fringe pattern.  We average the fringe pattern over all integrations per visit, and also over short and long fringe-periods, using a high-pass filter technique.  The high-pass filter smooths the visit-averaged and continuum-removed fringe pattern using a 10-point smoothing kernel.  Dividing the fringes by that smoothed version removes the long period portion of the fringe pattern, leaving the short period ($\sim$\,10 pixel) fringes.  The smoothed version is a good representation of the long period fringes.  This filtering produces two fringe basis vectors, short and long period fringes, that we fit to each individual spectrum as we now describe.

For a given individual spectrum, we remove its fringes by first dividing by a fitted polynomial to remove the continuum.  We then fit the resultant fringe pattern as a linear combination of the short- and long period basis vectors, modulated to vary in amplitude as a function of  wavelength (i.e., with rows on the detector).  That modulation is accomplished by multiplying the fringe basis vectors by polynomials, and the coefficients of the polynomials are varied by a gradient expansion algorithm to achieve a nonlinear least squares fit to the observed fringes.  We consulted a Bayesian Information Criterion (BIC, \citealp{schwarz_1978}) to determine the order of the polynomials that give the best results without overfitting.  The BIC values had a broad minimum for polynomial orders 4 to 6, and we used order 4 for all three polynomials (continuum removal, and modulation of two basis vectors).  (For consistency, we keep order 4 for all spectra in all channels and both visits.)  Restricting the polynomial order to a minimum consistent with the BIC ensures that our fringe removal does not significantly attenuate real spectral features in the eclipse spectrum.  The lower panel of Figure~\ref{fig: fringes} illustrates a randomly selected spectrum before and after the fringe removal. More information on fringe removal and stability is given in \S\,\ref{sec: noise}.

\subsection{White Light Eclipse Curves}\label{sec: white_light}

\begin{figure}
\centering
\includegraphics[width=5in]{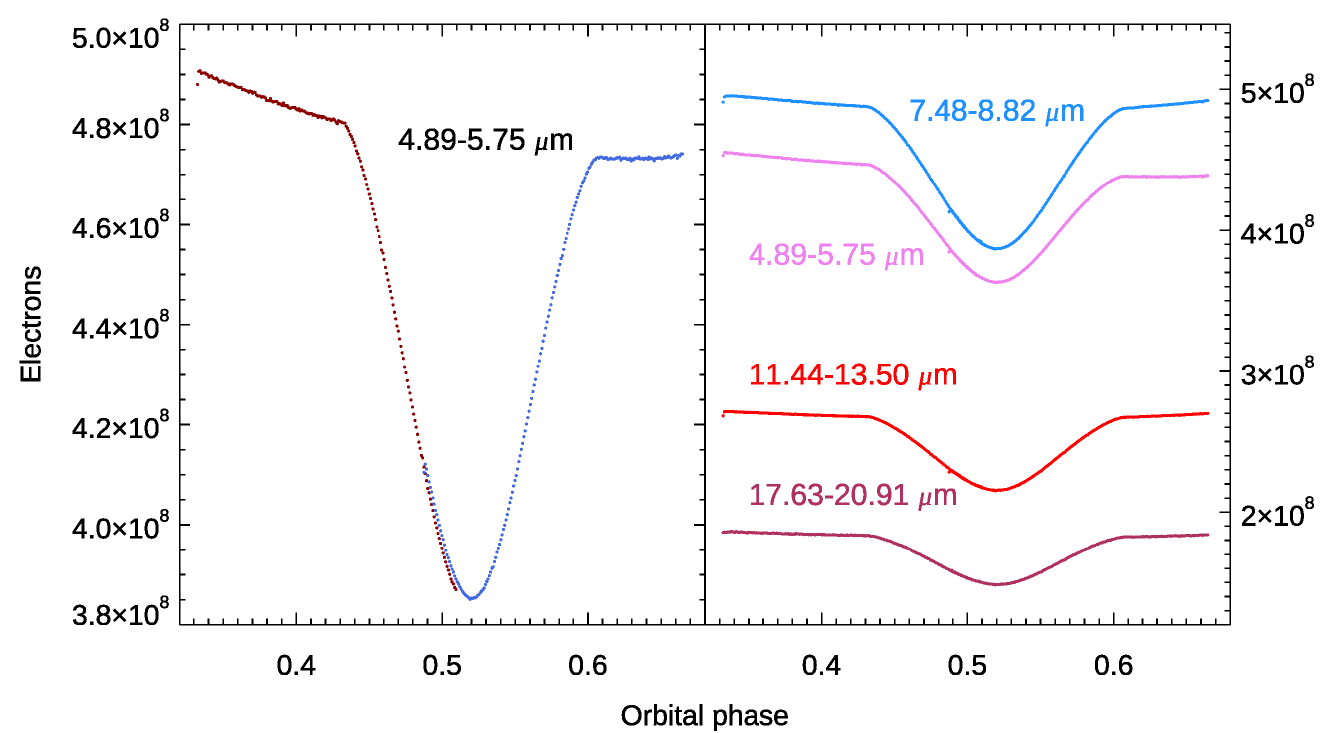}
\caption{White light eclipse curves for the 4 wavelength channels of R~CMa, obtained by summing the observed spectra over all wavelengths in each channel.  {\it Left panel:} "As observed" curves for channel-1 (4.89-5.75 $\mu$m), with no normalization or adjustments of any kind.  {\it Right panel:} Curves for each channel, after multiplying the ingress curve times a factor to agree with each egress curve in the region of overlap.  Also, the curves for channel-3 (11.44-13.50 $\mu$m), and channel-4 (17.63-20.91 $\mu$m) have been multiplied by 1.5 and by 8.0 respectively, for clarity of illustration.} 
\label{fig: wlight}
\end{figure}

With fringes removed from all of the observed spectra, we sum the spectra over wavelength to produce "white light" eclipse curves for each channel.  Those curves are illustrated in Figure~\ref{fig: wlight}.  In the left panel of that Figure, we plot the wavelength-summed ingress and egress light curves for channel-1, without any normalizations or adjustments of any kind. They overlap in the orbital phase range from 0.488 to 0.509, and their fluxes differ there by less than 1\%, due to the excellent stability and pointing accuracy of JWST.  These white light curves have very low noise, due to the large number of photoelectrons and the photometric stability of JWST (noise properties of the spectra are discussed further in \S\,\ref{sec: noise}.) 

We use white light curves for each channel in our process of solving for an eclipse spectrum, described below in \S\,\ref{sec: solving}.  For each channel, we combine the ingress and egress curves by applying a multiplicative factor close to unity to the ingress curve to force agreement between the amplitudes of the eclipse in the ingress versus egress visits. That factor turns out to be statistically indistinguishable from unity (median value $1.000052\pm0.000118$), showing that the ingress and egress eclipse amplitudes are closely equal.  The combined eclipse curves are shown in the right panel of Figure~\ref{fig: wlight}.

\subsection{Solving for Eclipse Depth}\label{sec: solving}

\begin{figure}
\centering
\includegraphics[width=3in]{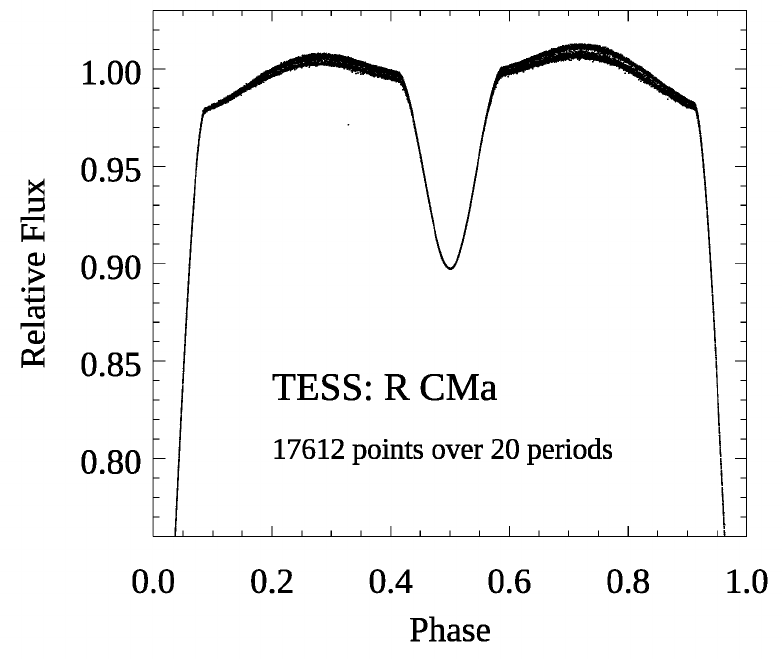}
\caption{Photometry of R~CMa's secondary eclipse based on TESS photometry over 20 orbital periods. The bandpass of TESS extends from 0.6- to 1.0\,$\mu$m, peaking at 0.787\,$\mu$m. }
\label{fig: tess}
\end{figure}

Our analysis to solve for an eclipse spectrum begins with the depths of the white light eclipses in each channel. Unlike most exoplanet eclipses, these data exhibit strong variation in flux outside of eclipse, due to the Roche lobe geometry of this semi-detached binary system.  We measure the white light eclipse depths by first normalizing each white light curve to unity in a region outside of eclipse.  That region is taken to span orbital phases 0.38-0.42 prior to ingress, and 0.62-0.66 after egress.  Those numerical ranges are arbitrary to some degree, because the system does not have a constant flux outside of eclipse.  Hence, we constructed the full light curve of R~CMa using photometry from TESS (\citealp{ricker_2015}, Figure~\ref{fig: tess}), and thereby we determined that the peak light of the system (both stars contributing) is 0.7\% greater than our nominal normalizing flux.  We corrected our normalizing factor by that amount (1.007).  Hence, all of our eclipse depths are relative to the maximum total light of the system.

To measure the white light eclipse depth, we fit parabolas to the minima region of each eclipse curve, and the white light depth is equal to unity (from the normalization) minus the flux at the minimum of the fitted parabola.  We calculate uncertainties on the fitted eclipse depth and central phase using a Monte-Carlo bootstrap procedure.  For comparison, we also derive the secondary eclipse depth in the TESS photometry, phasing and averaging the TESS data over 20 orbital periods, shown in Figure~\ref{fig: tess}. The secondary eclipse in the TESS optical band has a depth of approximately 0.1, versus a depth of approximately 0.2 in the MRS channels. That difference is expected because the cool secondary star in the system \citep{lehmann_2018} produces a larger fraction of the total light with increasing wavelength.  Comparison of the white light eclipse depths from TESS and the MRS to model atmospheres is discussed in \S\,\ref{sec: broad_band}.

To obtain an eclipse spectrum using the MRS data, we implement a differential technique, that has precedent in transit spectroscopy of exoplanets \citep{deming_2013, knutson_2014, stevenson_2014}.  At each wavelength, we divide the monochromatic spectral flux by the flux summed over all wavelengths in that channel, i.e. we divide by the white light eclipse curve for that channel.  When the monochromatic eclipse is deeper or more shallow than the white light eclipse, the ratio will exhibit a residual eclipse whose positive or negative differential depth we measure using linear regression.  Adding the depths of those differential eclipses at each wavelength to the measured depth of the white light eclipse for that spectral channel yields the monochromatic transit depth at each wavelength, i.e. the eclipse spectrum.  (The same procedure could be used to measure a transit spectrum, had we observed at transit rather than eclipse.)

\begin{figure}
\centering
\includegraphics[width=3in]{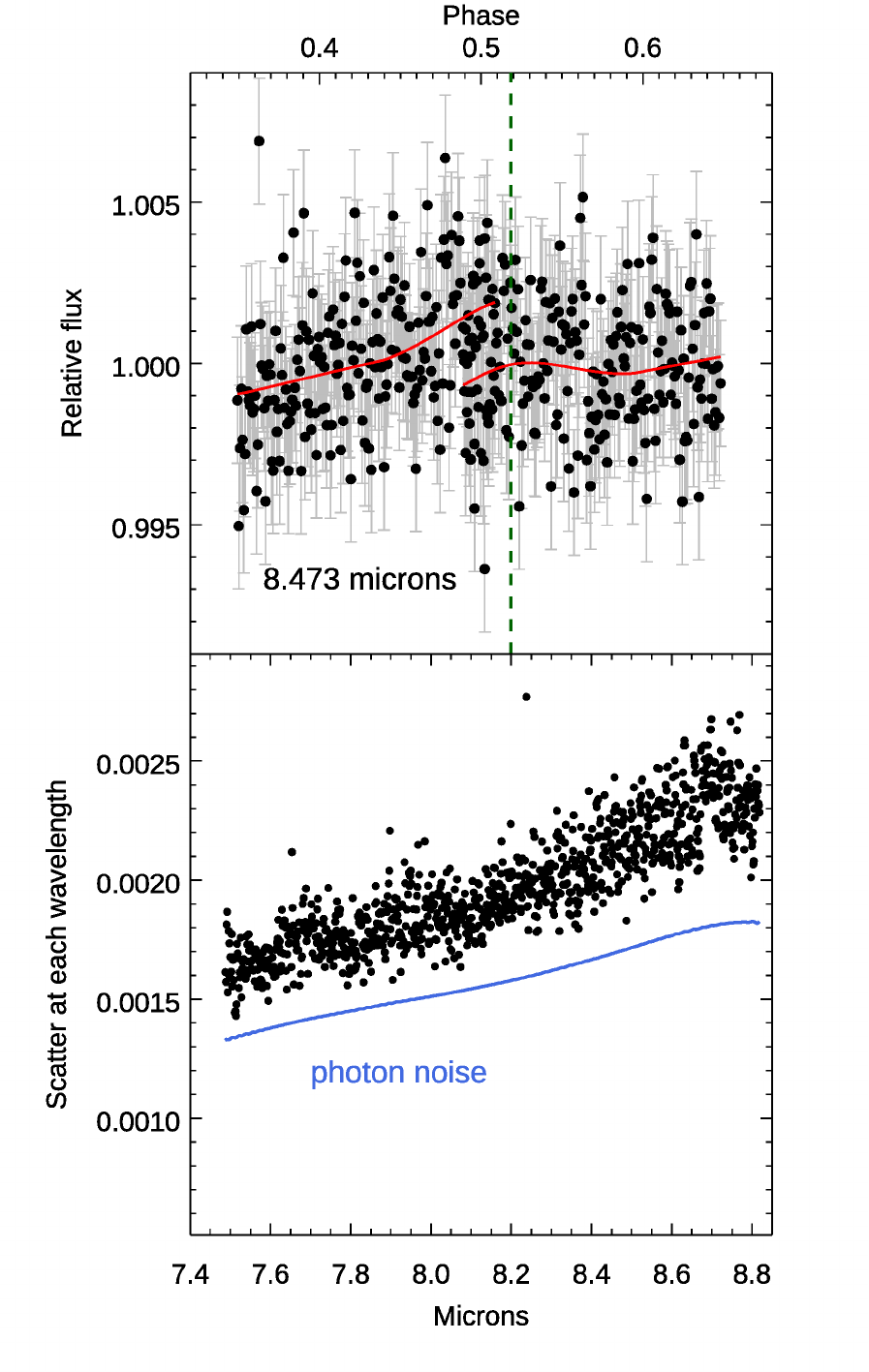}
\caption{{\it Upper panel:} Example of a differential eclipse fit for a randomly selected wavelength in channel-2.  Each point is a single integration at a single wavelength (row of the detector), plotted versus orbital phase (upper axis label), and error bars are shown in gray.  The red curves are the fitted ingress and egress portions of the white light eclipse curve for channel-2.  The dashed green line marks the fitted center of eclipse. The amplitude of the discontinuity between ingress and egress is one of the fitted parameters, and the ingress and egress amplitudes are constrained to be equal in the fitting process.  In this case the differential eclipse depth is negative, i.e. the monochromatic eclipse is not as deep as the white light eclipse. Note that the red curves are inverted, i.e. have negative amplitudes.  {\it Lower panel:} Noise level in the eclipse fitting versus wavelength (lower axis label) for all wavelengths in channel-2.  The noise level corresponds to the scatter in the data minus fit, as illustrated for one wavelength in the upper panel.  The observed noise is larger than the photon noise limit (blue curve) by a median factor of $1.24\pm0.04$.}
\label{fig: fit_example}
\end{figure}

Because the ingress and egress were observed in different visits, our regression solutions include an offset between the ingress and egress data that we derive as one of the fitted parameters.  We do not find it necessary to include any linear or higher order temporal baselines in the regression solution (e.g., \citealp{bouwman_2023}, such as could be produced by persistence and charge trapping \citep{morrison_2023}.   An example of fitting to a randomly selected differential eclipse is shown in Figure~\ref{fig: fit_example}. 

The method of fitting the white light curve to differential eclipses (called "divide-white" by \citealp{stevenson_2014}) has distinct advantages, especially in this case of R~CMa.  The white light eclipse curve of R~CMa is more complex than for exoplanets, and {\it ab initio} fitting would require consideration of effects such as strong rotational and tidal distortion, and both gravity and limb darkening \citep{wilson_1971}.  Modeling the TESS plus JWST light curves of this binary system using in-depth methods as per \citet{wilson_1971} would be quite valuable, but is beyond the scope of this paper.  The advantage of our differential technique is that it allows us to derive a good estimate of the eclipse spectra, largely independent of effects such as ellipticity and tidal distortion. The eclipse spectrum of R~CMa at the full resolution of the MRS is discussed in \S\,\ref{sec: hline}.

An aspect of eclipse spectroscopy that is untested by our work is to what degree spectra using different grating settings can be combined to construct a continuous spectrum, or whether offsets will appear between wavelength segments.  Such discontinuities are sometimes found in JWST's exoplanet transit spectroscopy \citealp{Fu_2024}.  We are unable to address that issue here because our current observations using only the 'A' grating setting.

\subsection{Comparison to the Pipeline}\label{sec: compare}

As explained in \S\,\ref{sec: extracting}, our analysis cannot fully use the standard calibration pipeline, because we combine pixels spatially at the group level in order to mitigate the BFE.  Nevertheless, we can compare two aspects of our analysis to results from the pipeline, namely linearity and defringing.

\subsubsection{A Linearity Comparison}\label{sec: linearity}
    
The ramp files that we use are corrected for non-linearity by the pipeline.  Nevertheless, we find it valuable to check for lingering effects of non-linearity in our results.  In \S\,\ref{sec: anal_ramp}, we found that the best signal-to-noise ratios in our spectra are obtained by subtracting the fitted ramp at the smallest group number from the fitted ramp at the largest group number.  That procedure precludes an explicit extraction of the linear term in the ramp solution, whereas the pipeline extracts that linear term when producing the rateints files.  We expect that both methods will yield similar results, especially at the longest wavelengths where the source flux is far from saturation.  Nevertheless, we compared our results to the 'rateints' files from the pipeline.  Zero-weighting the large number of 'NaN' pixels in the rateints files, we calculate white light eclipse curves, and compare those fitted depths to our analysis.  The only significant difference is for channel-1, the brightest portion of the spectrum.  In that instance, the eclipse depth from the pipeline is 2.7\% greater than our analysis (i.e., eclipse depth of $0.1966\pm0.0004$ versus $0.1915\pm0.0007$). In order to be consistent with conventional methodology, we use the pipeline value for the channel-1 depth in Table~\ref{table: depths}. We reiterate that we don't use the rateints files to produce spectra, because of the large number of 'NaN' pixels.

\subsubsection{A Defringing Comparison}~\label{sec: compare_defringe}

\begin{SCfigure}
\centering
\includegraphics[width=4in]{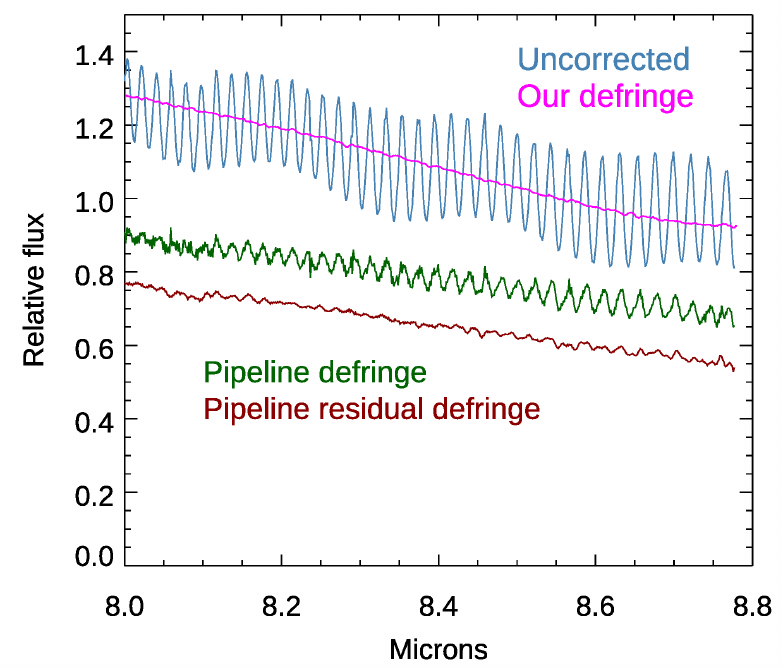}
\caption{Comparison of our defringing algorithm with defringing by the pipeline, using a portion of a randomly selected channel-2 spectrum.  This plots a wavelength range where the fringes are especially prominent.  The uncorrected spectrum at the top is overplotted with our defringed spectrum. The two stages of defringing by the pipeline are offset (for clarity) below the uncorrected spectrum. \label{fig: multiple_fringes} }
\end{SCfigure}

We have also compared our defringing algorithm to results from the pipeline, illustrated for one randomly selected channel-2 spectrum in Figure~\ref{fig: multiple_fringes} (we checked other spectra also, with similar results).  The pipeline fringe correction is a two-step process. First, each 2-D spectral frame is divided by a fringe-flat reference image, and then the remaining residual fringes are corrected using a 1-D fit of sinusoids to each spectrum\footnote{https://jwst-docs.stsci.edu/known-issues-with-jwst-data/miri-known-issues/miri-MRS-known-issues}, based on the known spatial frequencies of the fringes.  The pipeline fringe correction has very general applicability, suitable for many diverse science uses.  Our aggressive defringing method removes the fringes to lower amplitudes than does the pipeline (see Figure~\ref{fig: multiple_fringes}), but our method is specific to these particular data for R~CMa, and would not be appropriate for a range of different science cases. 

\subsection{Noise Properties}\label{sec: noise}

The spectra of transiting exoplanets have low amplitudes due to strong dilution by the flux from their host stars.  Therefore low noise observations are essential, and we have investigated the magnitude of the noise attained in our spectra of R~CMa, compared to the photon noise limit.  By photon noise limit, we mean the Poisson fluctuations in the total number of detected photons, and including the corresponding fluctuations in the background flux.  We neglect other noise sources such as detector read noise, because it is negligible compared to the photon noise of this bright source.  

The observed noise level at each wavelength is calculated as the standard deviation of the data minus the best fitting differential eclipse curve.  Figure~\ref{fig: fit_example} shows the scatter at one randomly selected wavelength in channel-2 (top panel), and the standard deviation plotted versus all channel-2 wavelengths and compared to the photon noise in the lower panel.  We achieve noise levels, expressed as ratios to the photon noise, of $1.24\pm0.03$, $1.24\pm0.04$, $0.99\pm0.09$, and $1.48\pm0.01$, in channels-1 through 4, respectively.  We also investigated how the residuals from the fitted monochromatic eclipse curves (\S\,\ref{sec: solving}) decrease when binned over $N$ temporal points. That behavior is called the Allan deviation relation \citep{allan_1966}.  Random white noise should produced a slope of 
$\frac{\Delta{log(\sigma_{N})}}{\Delta{log(N)}} = -0.5$, but red noise in the data would produce a slope that is less negative.  Median values of the Allan slopes for each wavelength channel are listed in Table~\ref{table: allan}, and they are close to the theoretical limiting value.

We consider the above noise properties to be excellent, with favorable implications for exoplanet spectroscopy, as we discuss in \S\,\ref{sec: planet_implications}.  Finally, we compared the noise level we achieved with the prediction of the Exposure Time Calculator (ETC) for the MRS.  For channel-2, our achieved noise level is 28\% less than predicted by the ETC, that we conclude is giving conservative estimates for the noise, at least in this case of R~CMa.

\begin{table}[h]
\centering
\begin{tabular}{lll}
 Wavelengths ($\mu$m)   &  Ingress  &  Egress \\
\hline
\hline
5.20-5.75 & $-0.522\pm0.009$ &    $-0.525\pm0.010$   \\
7.47-8.77 & $-0.496\pm0.004$  &    $-0.490\pm0.006$   \\
11.49-13.53 & $-0.432\pm0.006$  &   $-0.454\pm0.002$  \\
17.63-20.91 & $-0.498\pm0.006$  &    $-0.513\pm0.005$  \\
\end{tabular}
\caption{Allan deviation slopes \citep{allan_1966} for each channel, and the ingress and egress visits.  The Allan slopes express how the residuals in the fitted eclipse data vary with binning over time (see text). These slopes are median values over all wavelengths in each channel.}
\label{table: allan}
\end{table}

\begin{figure}
\centering
\includegraphics[width=4in]{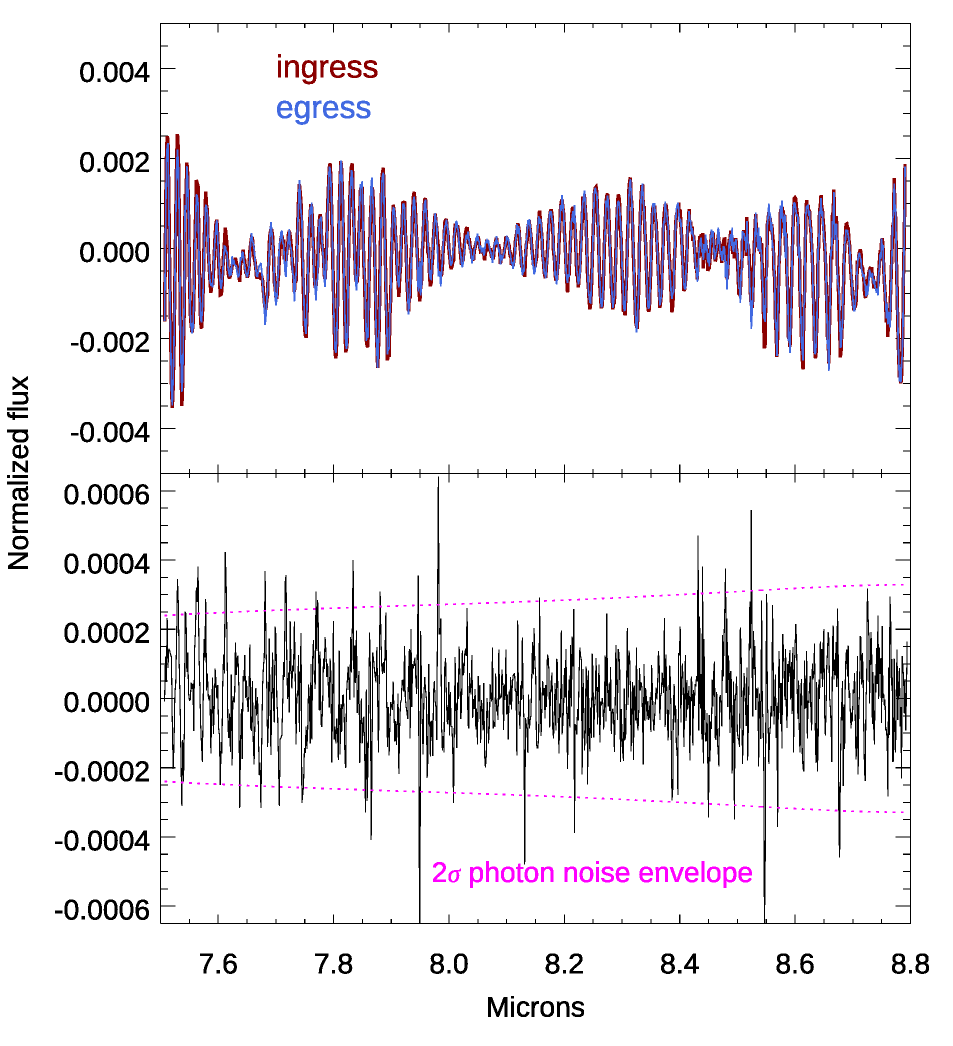}
\caption{{\it Top panel:} residual fringes remaining in our defringed spectra of R~CMa. The spectra are summed over each visit (203 spectra per visit), and the residual fringes are extracted using a high-pass filter.  The amplitude of these residual fringes (RMS\,$\sim$\,0.1\%) is more than an order of magnitude less than the fringes in the spectra before our defringing algorithm is applied (compare to Figure~\ref{fig: fringes}). The residual fringes in the ingress versus egress visits are nearly identical - the two sets of residual fringes virtually overplot. {\it Bottom panel:} Difference between the ingress and egress residual fringes.  The magenta lines are plotted at twice the theoretical photon noise envelope. } 
\label{fig: resid_fringes}
\end{figure}

In addition to the noise levels investigated above, we did an additional test related to the effectiveness of our defringing.  If the fringes vary in time, they could add noise to our eclipse spectra.  Although we regard our defringing method as excellent, it isn't perfect and there are very low amplitude residual fringes in our spectra. If the residual fringes vary during a single visit (thus affecting our eclipse depth solutions), then they would vary even more strongly from one visit to the next, because the star is re-acquired and occupies a slightly different position on the detector.  To investigate that possibility, we summed our defringed spectra over all integrations in each visit (ingress and egress).  Then we apply a high-pass filter to each of those summed spectra by dividing by a smoothing function, and that reveals the residual fringes.  

Figure~\ref{fig: resid_fringes} shows the residual fringes at ingress and egress, and they are virtually identical (top panel of the Figure).  The difference between them (bottom panel of the Figure) has a standard deviation of $1.38 \times 10^{-4}$ (138 ppm), in good agreement with the photon noise that varies with wavelength (flux level) from 118 to 164 ppm.  We conclude that, although our defringing method is not perfect, the residual fringes are stable and allow nearly photon-limited performance for exoplanet transits.  That excellent performance is enabled by JWST's exceptional pointing accuracy and stability.

\section{Results and Implications}\label{sec: results}

Our results have implications both for future observations of exoplanets in combined light (star+planet), and also for the secondary eclipse of R~CMa.

\subsection{Implications for Exoplanets}\label{sec: planet_implications}

Our results have implications both for transiting exoplanets, and also for close-in non-transiting worlds such as Promima Centauri\,b \citep{anglada_2016} whose atmosphere could potentially be characterized using the MRS \citep{snellen_2017}.

\subsubsection{Transits and Eclipses}\label{sec: transits}

There are many transiting exoplanets hosted by bright stars, wherein it would be problematic to use lower resolution spectroscopy with JWST because of saturation. The MRS provides an excellent alternative at wavelengths exceeding 4.88\,$\mu$m (\S\,\ref{sec: advantages}).  We conclude that optimal weighting of pixels at the group level (\S\,\ref{sec: anal_groups}), combined with a customized analysis of the detector ramp (\S\,\ref{sec: anal_ramp}), allows the MRS to achieve transit spectroscopy with noise close to the theoretical limit (\S\,\ref{sec: noise}).  Based on our experience observing R~CMa, we find that predictions of signal-to-noise ratio using JWST/MIRI's Exposure Time Calculator are conservative for bright sources observed with the MRS, not optimistic (\S\,\ref{sec: noise}).  Overall, our results support feasibility of exoplanet transit spectroscopy using the MRS, especially for bright systems that produce sufficient flux to enable using its full spectral resolving power.

\subsubsection{The Temperate Planet Orbiting Proxima Centauri}\label{sec: proxima}

Techniques such as transits, eclipses, and cross-correlation analyses, characterize exoplanetary atmospheres using the combined light of the planet and host star.  A fundamental principle of such techniques is that the exoplanet signal is easiest to detect when it is modulated in some fashion.  For example in-transit versus out-of-transit ratios, and/or exploiting large Doppler shifts of the exoplanet's spectrum, can isolate the exoplanet signal in the presence of instrumental effects and large flux from the host star.  Both \cite{kreidberg_2016} and \citet{snellen_2017} proposed using MIRI (the LRS, and the MRS, respectively) to detect the temperate planet orbiting Proxima Centauri \citep{anglada_2016}.  That planet does not transit \citep{jenkins_2019, gilbert_2021}, and its Doppler velocity would not be fully resolved by the MRS \citep{anglada_2016}.  

Although Proxima Centauri\,b does not transit, there are two methods that could potentially be used by JWST to characterize its atmosphere using the MIRI instrument. \citet{kreidberg_2016} propose measuring the phase curve at a wavelength corresponding to the peak of the planetary thermal emission, using MIRI's LRS to achieve a high rate of photon collection.  Since we have not studied data from the LRS in this paper, we cannot comment directly on the feasibility of that method.  However, \citet{bouwman_2023} and \citet{dyrek_2024a} found excellent performance by the LRS in time series observations, consistent with the close to photon-limited results we find here for the MRS.  

\citet{snellen_2017} propose that carbon dioxide at 15 $\mu$m could be detected in Proxima Centauri\,b using a spectral high-pass filter technique applied to spectra from the MRS.  A high pass filter would remove the continuum in the spectrum, and would detect CO$_2$ lines collectively (averaging over lines).  Nevertheless, this technique is challenging because instrumental effects (e.g., fringing) could interfere with the detection if they are not stable, given that they would not be subtracted via a temporal modulation such as a transit. 

Our results support the feasibility of the technique proposed by \cite{snellen_2017}, for two reasons.  First, we attain random noise levels in our extracted spectra that are close to the photon noise limit (\S\,\ref{sec: noise}). Second, we find that the instrumental fringes are stable over the 18-day interval between our two visits.  \citet{snellen_2017} projected that the instrumental spectral response (affected by fringing) would have to be stable to better than 10$^{-4}$ over tens of hours.  We find stability at $1.38 \times 10^{-4}$ level ($1-\sigma$, Figure~\ref{fig: resid_fringes}) over the 18 days between our two visits.  While that is not {\it less} than the 10$^{-4}$ level, the total integration time of our observations is less than Snellen et al. propose.  Our data do not cover the carbon dioxide region near 15\,$\mu$m, where the fringe properties may be different. Nevertheless, we believe that the method proposed by \citet{snellen_2017} is feasible based on a reasonable projection from the stability we observe with these R~CMa data near 8\,$\mu$m.

\subsection{Implications for R~Canis~Majoris}\label{sec: RCMa_implications}

Our results for R~CMa are twofold: 1) spectrophotometry in broad bands using the white light eclipse depths, and 2) the eclipse spectrum at the full spectral resolving power of the MRS.

\subsubsection{The Broadband Eclipse Spectrum}\label{sec: broad_band}

\begin{figure}
\centering
\includegraphics[width=4in]{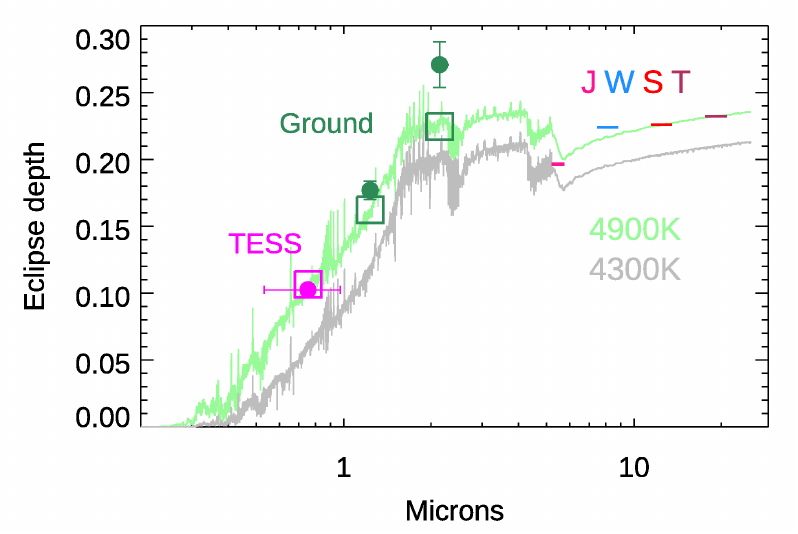}
\caption{Broadband secondary eclipse depths for R~CMa versus wavelength, compared to theoretical depths from model atmospheres at 4300 and 4900\,Kelvins. The depths from JWST's MRS and TESS are from this paper, and the ground-based depths are from \citet{varricatt_1999}. The solid points are observations and the open symbols are based on integrating Phoenix model atmospheric fluxes over the bandpass functions.  The small observed eclipse depth uncertainties for JWST are comparable to the widths of the lines that define the wavelength spans of the four channels.} 
\label{fig: broadband}
\end{figure}

The broadband eclipses in the four channels of the MRS give strong leverage on the temperature of the secondary star in the R~CMa system.  We compared the eclipse depth in each of those four bands (summing all wavelengths in each channel), plus the eclipse in the TESS band, and in the J- and K-band from ground-based photometry by \citet{varricatt_1999}.  Figure~\ref{fig: broadband} illustrates that comparison using Phoenix NextGen model atmospheres \citep{hauschildt_1999}, and adopting stellar radii and the primary star temperature (T=7300\,K) from \citet{budding_2011}.  We have not attempted a rigorous fit for the temperature of the secondary star, because that would require an analysis beyond our current scope, including tidal distortion and other proximity effects \citep{wilson_1971}.  Nevertheless, we obtain considerable insight into the temperature of the secondary based on the precision of the JWST eclipses and the simplicity of model atmospheres at long wavelength (minimal atomic and molecular absorptions).  Figure~\ref{fig: broadband} uses T=4900\,K for the secondary star, well above the T=4300\,K derived by \citet{budding_2011}.  Using T=4300\,K for the secondary star is incompatible with the eclipse depths that we measure using TESS and JWST.  We tried using stellar radii and a primary star temperature (T=7500\,K) from \cite{tomkin_1985} and \citet{tomkin_1989}, and those choices also require that the secondary star be significantly hotter than 4300\,K to be consistent with the TESS and JWST eclipses. \citet{varricatt_1999} also derived T=5158$\pm$47\,K from their analysis using K-band photometry. They pointed out that the temperature for the secondary star systematically increases with the wavelength of the optical-to-near-IR data used in previous analyses, and our JWST results are consistent with that trend. The atmospheric continuous opacity of FGK stars, due to H-minus, increases with wavelength in the infrared \citep{vernazza_1976}. Hence, increasing wavelengths are probing higher in the atmosphere of the secondary star. The highest layers of the secondary are likely to be heated by irradiation from the primary star \citep{claret_1992}, and that would qualitatively account for our results.

Table~\ref{table: depths} lists our derived white light eclipse depths in the four channels, and the central time of the eclipse (BJD-TDB) weighted over all 4 channels.  The high signal-to-noise of these JWST data enable a precision on the eclipse time of 15~seconds.

\begin{table}[h]
    \centering
    \begin{tabular}{ll}
     Wavelengths ($\mu$m)   &  Eclipse depth/time \\
      \hline
      \hline
        5.20-5.75 & $0.1966\pm0.0004$ \\
        7.47-8.77 & $0.2241\pm0.0001$ \\
        11.49-13.53 & $0.2261\pm0.0003$ \\
        17.63-20.91 & $0.2323\pm0.0003$ \\
        Central time & $2460019.44847\pm0.00017$ \\
    \end{tabular}
    \caption{White light secondary eclipse depths for R~CMa, summing the flux at all wavelengths in each channel of the MRS. The eclipse center time (in BJD/TDB) is a weighting over all four channels, and they are consistent.}
    \label{table: depths}
\end{table}

\subsubsection{Hydrogen Recombination Lines}\label{sec: hline}

\begin{figure}
 \plottwo{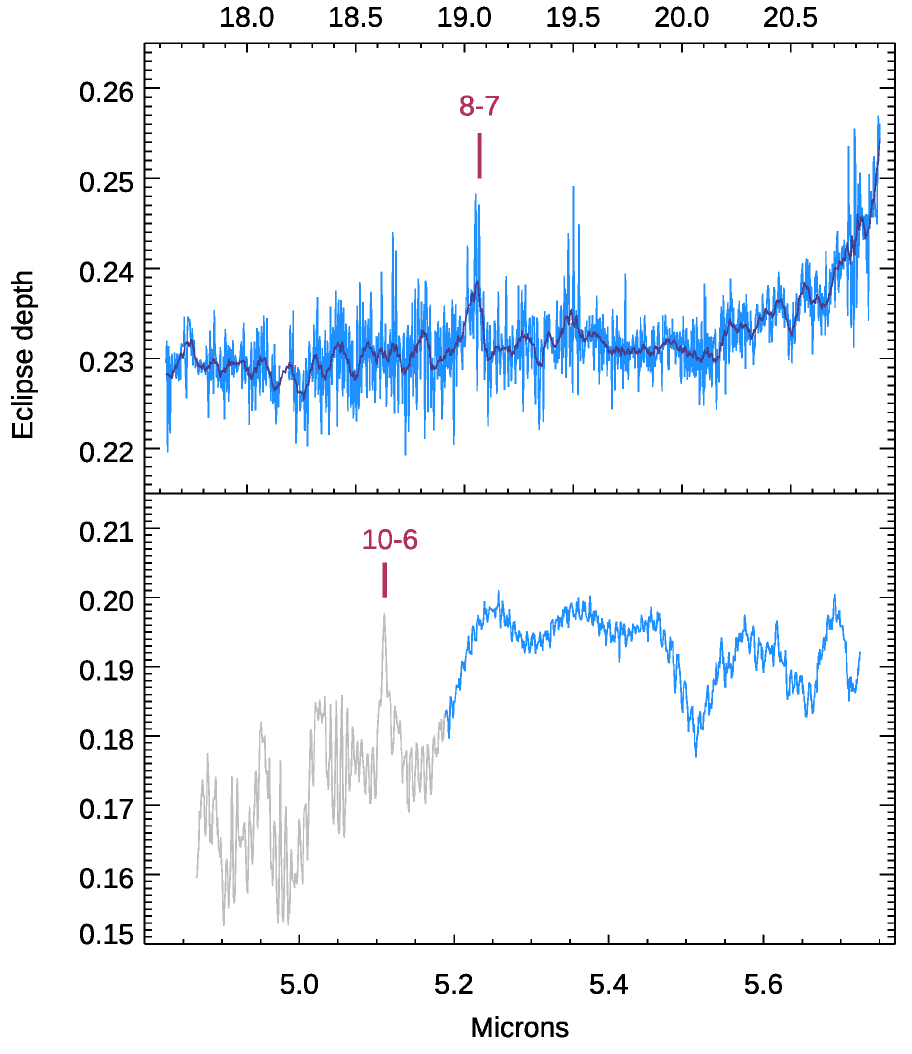}{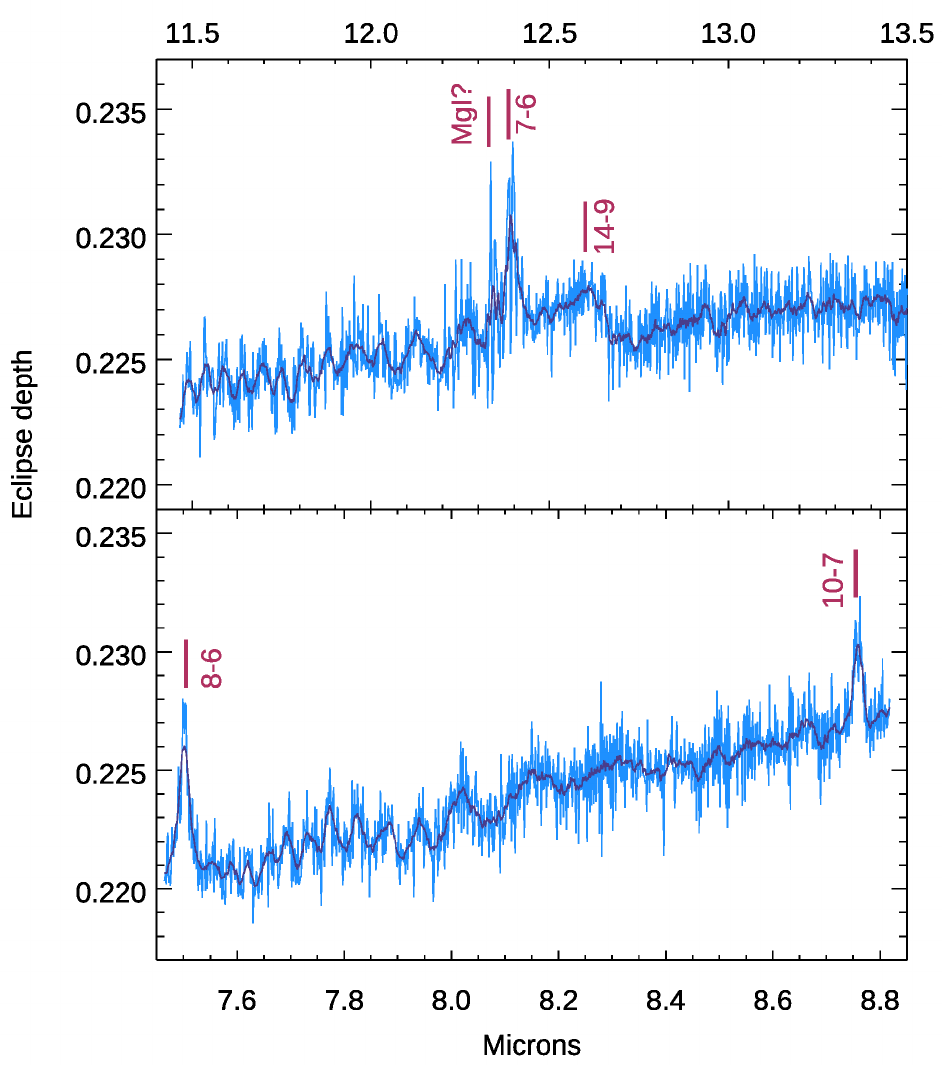}
\caption{Left: Secondary eclipse spectrum of R~CMa in channel-1 (bottom) and channel-4 (top).  The portion of the channel-1 spectrum plotted in gray is deemed unreliable due to saturation (see~\S\,\ref{sec: saturation}). Right: Eclipse spectrum in channels 2 (bottom) and 3 (top).  All spectra are plotted at single-pixel resolution (light blue), and smoothed over 10-pixels (dark blue).  Several hydrogen recombination emission lines are marked, with their upper and lower levels noted.  For example, the n=7 to n=6 line of atomic hydrogen is seen prominently at 12.37\,$\mu$m.  See text for discussion of the spectral features.}
\label{fig: spectra_hydrogen}
\end{figure}

Figure~\ref{fig: spectra_hydrogen} shows our eclipse spectra in each channel.  We group channels 1 and 4 in the same plot (left), because they are not as informative as channels-2 and -3 (on the right).  The portion of the channel-1 spectrum at wavelengths less than 5.2\,$\mu$m we deem to be unreliable due to saturation (\S\,\ref{sec: saturation}), and that portion is plotted in light gray on Figure~\ref{fig: spectra_hydrogen}.  

Collectively, the spectra show unequivocal evidence for atomic hydrogen emission, not seen previously in the Balmer lines \citep{lehmann_2018}.  All of the emission lines are weak, less than 1\% of the continuum flux from the system at each wavelength.  The statistical significance of their detection varies from line to line, but the overall appearance of very noticeable emission spikes in the data at wavelengths closely corresponding to atomic hydrogen transitions leaves no doubt that recombination emission is present.  Beginning in channel-1, there is a feature at 5.13\,$\mu$m that coincides with the $n=10$ to $n=6$ (10-6) transition, albeit we do not claim it because it lies in the saturated region.  However, in channel-2 the 8-6 and 10-7 transitions are clearly detected at 7.51 and 8.76\,$\mu$m, respectively.  In channel-3, the 12.0-12.4\,$\mu$m spectral region is affected by a second-order light leak from 6\,$\mu$m that is a potential source of confusion \citep{gasman_2023}. Nevertheless, the 7-6 line appears clearly at 12.37\,$\mu$m, and good evidence for the 14-9 transition at 12.6\,$\mu$m.  In channel-4, the 8-7 transition appears at 19.06\,$\mu$m.  

In all panels of Figure~\ref{fig: spectra_hydrogen}, we plot the eclipse spectra at full resolution (in light blue) and also smoothed over 10 wavelength points (in dark blue).  The smoothed versions of the spectra are very useful to indicate which features are real emission lines and which are noise.  For example, in the channel-4 spectrum there are two possible emission lines at 19.06 and also 19.50\,$\mu$m.  However, in the smoothed spectrum only the 19.06\,$\mu$m feature is prominent.  When a candidate "line" is merely a noisy region of the spectrum, the fluctuations in closely adjacent wavelengths tend to be inconsistent and tend to cancel in the smoothed spectrum, whereas real emission lines tend to reinforce. 

In the channel-3 spectrum there is a suggestion that the 7-6 transition in neutral Magnesium may appear at 12.32\,$\mu$m (Figure~\ref{fig: spectra_hydrogen}).  It is too weak to claim a detection, but it deserves future follow-up in some way (e.g., with ELTs).  This hydrogenic transition in Magnesium is prominent in the spectrum of the Sun \citep{brault_1983} and solar-type stars \citep{carlsson_1992}, and is {\it extremely} sensitive to magnetic fields \citep{deming_1988}.  It represents a potentially very useful probe of magnetic physics in the mass transfer process of Algol-type binaries, if it can be confirmed by future observations.

We attempted to derive the electron temperature and density that would be consistent with the hydrogen recombination lines that we detect, by comparing to theoretical line intensities versus temperature and density from \citet{storey_1995}.  However, it quickly became apparent that no single temperature can account for the observed line intensities.  The spectra in Figure~\ref{fig: spectra_hydrogen} are normalized to the total light of the system at each wavelength, and the total light decreases substantially with wavelength.  For example, the 8-6 and 7-6 line intensities (in photons per second) are predicted to be very similar at plausible temperatures, because their upper state energies do not differ greatly.  Indeed, their {\it normalized} emission line amplitudes are similar (Figure~\ref{fig: spectra_hydrogen}).  However, at the wavelength of the 8-6 transition the normalizing flux is more than 7 times larger than at the wavelength of the 7-6 line.  Consequently, the observed strengths of those two lines (in photons per second, before normalizing to the continuum) are very discrepant compared to theoretical values at any single plausible temperature and density.  We conclude that these recombination lines are sensing a range of temperature and optical depths. Heating of the outer layers of the secondary star by irradiation (noted in \S\,\ref{sec: broad_band}) would facilitate the excitation of these lines.

\section{Summary}\label{sec: summary}

In this paper we analyze MIRI/MRS time series observations of the secondary eclipse in the bright stellar eclipsing binary system R~Canis Majoris, using a single grating setting, as a test and validation of using the MRS for transit spectroscopy of exoplanets.  The MRS has some important advantages for transit and eclipse spectroscopy of exoplanets, explained in \S\,\ref{sec: advantages}. R~CMa is a bright source (V=5.7), and due to data volume limitations in APT we observed it in two visits, separately covering ingress and egress (\S\,\ref{sec: observations}). Each of 203 integrations per visit used 30 groups to read the detectors, producing a large total electron charge.  This large charge accumulation causes discontinuous and large (exceeding hundreds of DNs) negative jumps in the detector ramps for individual pixels (\S\,\ref{sec: jumps}).  We performed a statistical analysis of the jumps (\S\,\ref{sec: stat_jumps}) and we find that they tend to occur when pixels are close to saturation, but not always, and not in a 1-to-1 relation with saturated pixels. We concluded  that the negative jumps are produced by the pipeline when it makes the ramp files, as explained in \S\,\ref{sec: stat_jumps}. We have developed a custom pipeline to analyze these data, including correcting the jumps using a running median filter applied to the time differences in the detector ramps (\S\,\ref{sec: anal_ramp}).  

The strong spatial contrast in flux for this bright star produces a prominent brighter-fatter effect (BFE) for pixels adjacent to the brightest portions of the spectrum.  To deal with the BFE, we perform an optimal extraction at the group level (\S\,\ref{sec: anal_groups}) that helps to recapture charge that spatially migrates between pixels.  The sequence of extracted fluxes at the group level (\S\,\ref{sec: extracting}) produces a ramp function that requires a 3rd or 4th order polynomial to account for slope variations in the ramp (\S\,\ref{sec: anal_ramp}).  There is saturation at the shortest wavelengths, that we partially correct by restricting the range of groups used in the ramp.  Nevertheless, extracted spectra at wavelengths below 5.2\,$\mu$m are not reliable due to saturation (\S\,\ref{sec: saturation}).  However, every star hosting a transiting planet can be observed using charge levels on the detector that are less than we use here. 

The MRS is an excellent choice for observing a wide range of exoplanets transiting very bright stars, because:

\begin{itemize}
    \item { The MRS is JWST's highest spectral resolution mode, reaching to the longest wavelength compared to other instruments (\S\ref{sec: MRS}).}
    \item {Observations using the MRS for transiting planet systems can avoid saturation over most of its wavelength range, for even the brightest planet-hosting stars (\S\ref{sec: saturation}).}
    \item{We demonstrate nearly photon-limited signal-to-noise ratios (\S\ref{sec: noise}), that increase close to a square-root dependence when binning in time (Table~\ref{table: allan}). }
    \item{ Spectra from the MRS have fringes, but our custom pipeline includes a defringing algorithm (\S\,\ref{sec: defringe}) that reduces the fringes in the spectra to $\sim\,0.1\%$ amplitude.  Due to JWST's excellent pointing accuracy and stability, the remaining residual fringes are stable to at least 138 parts per million (limited by our photon noise) over the 18 days between our ingress and egress visits (\S\,\ref{sec: planet_implications}). The fringe stability supports the viability of the method proposed by \citet{snellen_2017} to detect a carbon dioxide atmosphere on Proxima Cen\,b using the MRS (\S\,\ref{sec: planet_implications}). }
    \item{Our observations use only the 'A/SHORT' grating setting, and we cannot determine to what degree exoplanet transit spectroscopy might encounter offsets in transit spectra  between wavelength segments of different grating settings (\S\ref{sec: solving}).  Testing that issue would be a valuable extension of our work. }
\end{itemize}

Because we measure the secondary eclipse of R~CMa with high signal-to-noise ratio, we derive new scientific information on the secondary star in this Algol-type binary.  Already it was known that the secondary star has expanded to fill its Roche lobe, and is likely to be transferring mass to the primary star.  Previous observations have noted that the effective temperature required to match observations of the secondary star becomes greater at increasingly long wavelengths.  We measured the depth of the secondary eclipse in photometry from TESS, and in wavelength-integrated (white light) bands by summing the the MRS data over the wavelengths in each of our 4 channels (\S\,\ref{sec: solving} and Table~\ref{table: depths}).  Comparing to Phoenix model atmospheres, we find that the TESS+JWST/MRS data require a temperature for the secondary star that is significantly hotter (T$\sim$4900\,K) than the $T=4300$\,K temperature derived by optical observers.  We suggest that conventional stellar model atmospheres do not adequately account for the layered structure of a Roche-lobe filling star of this type, probably due to irradiation by the primary star.  We also measure a precise time for the center of secondary eclipse (Table~\ref{table: depths}), that can be used to improve the orbital ephemeris of this system.

Our spectroscopy using the MRS shows atomic hydrogen emission lines from the secondary star (\S\,\ref{sec: hline}), from upper levels $n=$7, 8, 10 and 14, probably populated by recombination. The lines are weak, each line being less than 1\% of the continuum from the combined light of both stars.  Nevertheless, the lines are unequivocally detected in emission in these low-noise spectra. Comparing to theoretical recombination calculations, we find that no single electron temperature and density can account for the relative line intensities.  We conclude that the emission lines probe a range of continuum optical depths in the layered atmosphere of the extended secondary star.

\acknowledgements

We thank the referee for a host of insightful comments that improved this paper, and we (especially D.D.) thank the Help Desk staff at the Space Telescope Science Institute for their patient explanations and answers to our many questions. This work is based in part on observations made with the NASA/ESA/CSA James Webb Space Telescope. The data were obtained from the Mikulski Archive for Space Telescopes at the Space Telescope Science Institute, which is operated by the Association of Universities for Research in Astronomy, Inc., under NASA contract NAS 5-03127 for JWST. These observations are associated with program JWST-GO-01556.   Support for program JWST-GO-01556 was provided by NASA through a grant from the Space Telescope Science Institute to the University of Maryland.

% \begin{thebibliography}{references.bib}

\clearpage

\bibliography{references.bib}

\begin{thebibliography}{}
\expandafter\ifx\csname natexlab\endcsname\relax\def\natexlab#1{#1}\fi
\providecommand{\url}[1]{\href{#1}{#1}}
\providecommand{\dodoi}[1]{doi:~\href{http://doi.org/#1}{\nolinkurl{#1}}}
\providecommand{\doeprint}[1]{\href{http://ascl.net/#1}{\nolinkurl{http://ascl.net/#1}}}
\providecommand{\doarXiv}[1]{\href{https://arxiv.org/abs/#1}{\nolinkurl{https://arxiv.org/abs/#1}}}

\bibitem[{{Ahrer} {et~al.}(2023){Ahrer}, {Stevenson}, {Mansfield}, {Moran},
  {Brande}, {Morello}, {Murray}, {Nikolov}, {Petit dit de la Roche},
  {Schlawin}, {Wheatley}, {Zieba}, {Batalha}, {Damiano}, {Goyal}, {Lendl},
  {Lothringer}, {Mukherjee}, {Ohno}, {Batalha}, {Battley}, {Bean}, {Beatty},
  {Benneke}, {Berta-Thompson}, {Carter}, {Cubillos}, {Daylan}, {Espinoza},
  {Gao}, {Gibson}, {Gill}, {Harrington}, {Hu}, {Kreidberg}, {Lewis}, {Line},
  {L{\'o}pez-Morales}, {Parmentier}, {Powell}, {Sing}, {Tsai}, {Wakeford},
  {Welbanks}, {Alam}, {Alderson}, {Allen}, {Anderson}, {Barstow}, {Bayliss},
  {Bell}, {Blecic}, {Bryant}, {Burleigh}, {Carone}, {Casewell}, {Changeat},
  {Chubb}, {Crossfield}, {Crouzet}, {Decin}, {D{\'e}sert}, {Feinstein},
  {Flagg}, {Fortney}, {Gizis}, {Heng}, {Iro}, {Kempton}, {Kendrew}, {Kirk},
  {Knutson}, {Komacek}, {Lagage}, {Leconte}, {Lustig-Yaeger}, {MacDonald},
  {Mancini}, {May}, {Mayne}, {Miguel}, {Mikal-Evans}, {Molaverdikhani},
  {Palle}, {Piaulet}, {Rackham}, {Redfield}, {Rogers}, {Roy}, {Rustamkulov},
  {Shkolnik}, {Sotzen}, {Taylor}, {Tremblin}, {Tucker}, {Turner}, {de
  Val-Borro}, {Venot}, \& {Zhang}}]{ahrer_2023}
{Ahrer}, E.-M., {Stevenson}, K.~B., {Mansfield}, M., {et~al.} 2023, Nature,
  614, 653, \dodoi{10.1038/s41586-022-05590-4}

\bibitem[{{Albert} {et~al.}(2023){Albert}, {Lafreni{\`e}re}, {Ren{\'e}},
  {Artigau}, {Volk}, {Goudfrooij}, {Martel}, {Radica}, {Rowe}, {Espinoza},
  {Roy}, {Filippazzo}, {Darveau-Bernier}, {Talens}, {Sivaramakrishnan},
  {Willott}, {Fullerton}, {LaMassa}, {Hutchings}, {Rowlands}, {Vila}, {Zhou},
  {Aldridge}, {Maszkiewicz}, {Beaulieu}, {Cook}, {Piaulet}, {Roy},
  {Lamontagne}, {Morel}, {Frost}, {Salhi}, {Coulombe}, {Benneke}, {MacDonald},
  {Johnstone}, {Turner}, {Fournier-Tondreau}, {Allart}, \&
  {Kaltenegger}}]{albert_2023}
{Albert}, L., {Lafreni{\`e}re}, D., {Ren{\'e}}, D., {et~al.} 2023, PASP, 135,
  075001, \dodoi{10.1088/1538-3873/acd7a3}

\bibitem[{{Alderson} {et~al.}(2023){Alderson}, {Wakeford}, {Alam}, {Batalha},
  {Lothringer}, {Adams Redai}, {Barat}, {Brande}, {Damiano}, {Daylan},
  {Espinoza}, {Flagg}, {Goyal}, {Grant}, {Hu}, {Inglis}, {Lee}, {Mikal-Evans},
  {Ramos-Rosado}, {Roy}, {Wallack}, {Batalha}, {Bean}, {Benneke},
  {Berta-Thompson}, {Carter}, {Changeat}, {Col{\'o}n}, {Crossfield},
  {D{\'e}sert}, {Foreman-Mackey}, {Gibson}, {Kreidberg}, {Line},
  {L{\'o}pez-Morales}, {Molaverdikhani}, {Moran}, {Morello}, {Moses},
  {Mukherjee}, {Schlawin}, {Sing}, {Stevenson}, {Taylor}, {Aggarwal}, {Ahrer},
  {Allen}, {Barstow}, {Bell}, {Blecic}, {Casewell}, {Chubb}, {Crouzet},
  {Cubillos}, {Decin}, {Feinstein}, {Fortney}, {Harrington}, {Heng}, {Iro},
  {Kempton}, {Kirk}, {Knutson}, {Krick}, {Leconte}, {Lendl}, {MacDonald},
  {Mancini}, {Mansfield}, {May}, {Mayne}, {Miguel}, {Nikolov}, {Ohno}, {Palle},
  {Parmentier}, {Petit dit de la Roche}, {Piaulet}, {Powell}, {Rackham},
  {Redfield}, {Rogers}, {Rustamkulov}, {Tan}, {Tremblin}, {Tsai}, {Turner}, {de
  Val-Borro}, {Venot}, {Welbanks}, {Wheatley}, \& {Zhang}}]{alderson_2023}
{Alderson}, L., {Wakeford}, H.~R., {Alam}, M.~K., {et~al.} 2023, Nature, 614,
  664, \dodoi{10.1038/s41586-022-05591-3}

\bibitem[{{Allan}(1966)}]{allan_1966}
{Allan}, D.~W. 1966, IEEE Proceedings, 54, 221, \dodoi{10.1109/PROC.1966.4634}

\bibitem[{{Anglada-Escud{\'e}} {et~al.}(2016){Anglada-Escud{\'e}}, {Amado},
  {Barnes}, {Berdi{\~n}as}, {Butler}, {Coleman}, {de La Cueva}, {Dreizler},
  {Endl}, {Giesers}, {Jeffers}, {Jenkins}, {Jones}, {Kiraga}, {K{\"u}rster},
  {L{\'o}pez-Gonz{\'a}lez}, {Marvin}, {Morales}, {Morin}, {Nelson}, {Ortiz},
  {Ofir}, {Paardekooper}, {Reiners}, {Rodr{\'\i}guez},
  {Rodr{\'\i}guez-L{\'o}pez}, {Sarmiento}, {Strachan}, {Tsapras}, {Tuomi}, \&
  {Zechmeister}}]{anglada_2016}
{Anglada-Escud{\'e}}, G., {Amado}, P.~J., {Barnes}, J., {et~al.} 2016, Nature,
  536, 437, \dodoi{10.1038/nature19106}

\bibitem[{Argyriou(2021)}]{argyriou_2021}
Argyriou, I. 2021, Phd thesis, Institute of Astronomy, KU Leuven, Leuven,
  Belgium

\bibitem[{{Argyriou} {et~al.}(2020{\natexlab{a}}){Argyriou}, {Rieke},
  {Ressler}, {G{\'a}sp{\'a}r}, \& {Vandenbussche}}]{argyriou_2020a}
{Argyriou}, I., {Rieke}, G.~H., {Ressler}, M.~E., {G{\'a}sp{\'a}r}, A., \&
  {Vandenbussche}, B. 2020{\natexlab{a}}, in Society of Photo-Optical
  Instrumentation Engineers (SPIE) Conference Series, Vol. 11454, X-Ray,
  Optical, and Infrared Detectors for Astronomy IX, ed. A.~D. {Holland} \&
  J.~{Beletic}, 114541P, \dodoi{10.1117/12.2561502}

\bibitem[{{Argyriou} {et~al.}(2020{\natexlab{b}}){Argyriou}, {Wells}, {Glasse},
  {Lee}, {Royer}, {Vandenbussche}, {Malumuth}, {Glauser}, {Kavanagh},
  {Labiano}, {Lahuis}, {Mueller}, \& {Patapis}}]{argyriou_2020b}
{Argyriou}, I., {Wells}, M., {Glasse}, A., {et~al.} 2020{\natexlab{b}},
  Astr.Ap., 641, A150, \dodoi{10.1051/0004-6361/202037535}

\bibitem[{{Argyriou} {et~al.}(2023){Argyriou}, {Glasse}, {Law}, {Labiano},
  {{\'A}lvarez-M{\'a}rquez}, {Patapis}, {Kavanagh}, {Gasman}, {Mueller},
  {Larson}, {Vandenbussche}, {Glauser}, {Royer}, {Dicken}, {Harkett},
  {Sargent}, {Engesser}, {Jones}, {Kendrew}, {Noriega-Crespo}, {Brandl},
  {Rieke}, {Wright}, {Lee}, \& {Wells}}]{argyriou_2023}
{Argyriou}, I., {Glasse}, A., {Law}, D.~R., {et~al.} 2023, Astr.Ap., 675, A111,
  \dodoi{10.1051/0004-6361/202346489}

\bibitem[{{Arulanantham} {et~al.}(2024){Arulanantham}, {McClure},
  {Pontoppidan}, {Beck}, {Sturm}, {Harsono}, {Boogert}, {Cordiner}, {Dartois},
  {Drozdovskaya}, {Espaillat}, {Melnick}, {Noble}, {Palumbo}, {Pendleton},
  {Terada}, \& {van Dishoeck}}]{arulanantham_2024}
{Arulanantham}, N., {McClure}, M.~K., {Pontoppidan}, K., {et~al.} 2024,
  ApJ(Lett), 965, L13, \dodoi{10.3847/2041-8213/ad35c9}

\bibitem[{{August} {et~al.}(2023){August}, {Bean}, {Zhang}, {Lunine}, {Xue},
  {Line}, \& {Smith}}]{august_2023}
{August}, P.~C., {Bean}, J.~L., {Zhang}, M., {et~al.} 2023, Ap.J.(Lett.), 953,
  L24, \dodoi{10.3847/2041-8213/ace828}

\bibitem[{{Batalha} {et~al.}(2018){Batalha}, {Lewis}, {Line}, {Valenti}, \&
  {Stevenson}}]{batalha_2018}
{Batalha}, N.~E., {Lewis}, N.~K., {Line}, M.~R., {Valenti}, J., \& {Stevenson},
  K. 2018, Ap.J.(Lett.), 856, L34, \dodoi{10.3847/2041-8213/aab896}

\bibitem[{{Bell} {et~al.}(2024){Bell}, {Crouzet}, {Cubillos}, {Kreidberg},
  {Piette}, {Roman}, {Barstow}, {Blecic}, {Carone}, {Coulombe}, {Ducrot},
  {Hammond}, {Mendon{\c{c}}a}, {Moses}, {Parmentier}, {Stevenson},
  {Teinturier}, {Zhang}, {Batalha}, {Bean}, {Benneke}, {Charnay}, {Chubb},
  {Demory}, {Gao}, {Lee}, {L{\'o}pez-Morales}, {Morello}, {Rauscher}, {Sing},
  {Tan}, {Venot}, {Wakeford}, {Aggarwal}, {Ahrer}, {Alam}, {Baeyens},
  {Barrado}, {Caceres}, {Carter}, {Casewell}, {Challener}, {Crossfield},
  {Decin}, {D{\'e}sert}, {Dobbs-Dixon}, {Dyrek}, {Espinoza}, {Feinstein},
  {Gibson}, {Harrington}, {Helling}, {Hu}, {Iro}, {Kempton}, {Kendrew},
  {Komacek}, {Krick}, {Lagage}, {Leconte}, {Lendl}, {Lewis}, {Lothringer},
  {Malsky}, {Mancini}, {Mansfield}, {Mayne}, {Evans-Soma}, {Molaverdikhani},
  {Nikolov}, {Nixon}, {Palle}, {Petit dit de la Roche}, {Piaulet}, {Powell},
  {Rackham}, {Schneider}, {Steinrueck}, {Taylor}, {Welbanks}, {Yurchenko},
  {Zhang}, \& {Zieba}}]{bell_2024}
{Bell}, T.~J., {Crouzet}, N., {Cubillos}, P.~E., {et~al.} 2024, Nature
  Astronomy, \dodoi{10.1038/s41550-024-02230-x}

\bibitem[{{Birkby} {et~al.}(2013){Birkby}, {de Kok}, {Brogi}, {de Mooij},
  {Schwarz}, {Albrecht}, \& {Snellen}}]{birkby_2013}
{Birkby}, J.~L., {de Kok}, R.~J., {Brogi}, M., {et~al.} 2013, MNRAS, 436, L35,
  \dodoi{10.1093/mnrasl/slt107}

\bibitem[{{Birkmann} {et~al.}(2022){Birkmann}, {Ferruit}, {Giardino},
  {Nielsen}, {Garc{\'\i}a Mu{\~n}oz}, {Kendrew}, {Rauscher}, {Beck}, {Keyes},
  {Valenti}, {Jakobsen}, {Dorner}, {Alves de Oliveira}, {Arribas}, {B{\"o}ker},
  {Bunker}, {Charlot}, {de Marchi}, {Kumari}, {L{\'o}pez-Caniego},
  {L{\"u}tzgendorf}, {Maiolino}, {Manjavacas}, {Marston}, {Moseley}, {Prizkal},
  {Proffitt}, {Rawle}, {Rix}, {te Plate}, {Sabbi}, {Sirianni}, {Willott}, \&
  {Zeidler}}]{birkmann_2022}
{Birkmann}, S.~M., {Ferruit}, P., {Giardino}, G., {et~al.} 2022, Astr.Ap., 661,
  A83, \dodoi{10.1051/0004-6361/202142592}

\bibitem[{{Bouwman} {et~al.}(2023){Bouwman}, {Kendrew}, {Greene}, {Bell},
  {Lagage}, {Schreiber}, {Dicken}, {Sloan}, {Espinoza}, {Scheithauer},
  {Coulais}, {Fox}, {Gastaud}, {Glauser}, {Jones}, {Labiano}, {Lahuis},
  {Morrison}, {Murray}, {Mueller}, {Nayak}, {Wright}, {Glasse}, \&
  {Rieke}}]{bouwman_2023}
{Bouwman}, J., {Kendrew}, S., {Greene}, T.~P., {et~al.} 2023, PASP, 135,
  038002, \dodoi{10.1088/1538-3873/acbc49}

\bibitem[{{Brault} \& {Noyes}(1983)}]{brault_1983}
{Brault}, J., \& {Noyes}, R. 1983, Ap.J.(Lett.), 269, L61,
  \dodoi{10.1086/184056}

\bibitem[{{Brogi} {et~al.}(2012){Brogi}, {Snellen}, {de Kok}, {Albrecht},
  {Birkby}, \& {de Mooij}}]{brogi_2012}
{Brogi}, M., {Snellen}, I. A.~G., {de Kok}, R.~J., {et~al.} 2012, Nature, 486,
  502, \dodoi{10.1038/nature11161}

\bibitem[{{Budding} \& {Butland}(2011)}]{budding_2011}
{Budding}, E., \& {Butland}, R. 2011, MNRAS, 418, 1764,
  \dodoi{10.1111/j.1365-2966.2011.19597.x}

\bibitem[{{Carlsson} {et~al.}(1992){Carlsson}, {Rutten}, \&
  {Shchukina}}]{carlsson_1992}
{Carlsson}, M., {Rutten}, R.~J., \& {Shchukina}, N.~G. 1992, Astr.Ap., 253, 567

\bibitem[{{Chandler}(1887)}]{chandler_1887}
{Chandler}, S.~C. 1887, Astron.J., 7, 150, \dodoi{10.1086/100957}

\bibitem[{{Claret} \& {Gimenez}(1992)}]{claret_1992}
{Claret}, A., \& {Gimenez}, A. 1992, Astr.Ap., 256, 572

\bibitem[{{Coulombe} {et~al.}(2023){Coulombe}, {Benneke}, {Challener},
  {Piette}, {Wiser}, {Mansfield}, {MacDonald}, {Beltz}, {Feinstein}, {Radica},
  {Savel}, {Dos Santos}, {Bean}, {Parmentier}, {Wong}, {Rauscher}, {Komacek},
  {Kempton}, {Tan}, {Hammond}, {Lewis}, {Line}, {Lee}, {Shivkumar},
  {Crossfield}, {Nixon}, {Rackham}, {Wakeford}, {Welbanks}, {Zhang}, {Batalha},
  {Berta-Thompson}, {Changeat}, {D{\'e}sert}, {Espinoza}, {Goyal},
  {Harrington}, {Knutson}, {Kreidberg}, {L{\'o}pez-Morales}, {Shporer}, {Sing},
  {Stevenson}, {Aggarwal}, {Ahrer}, {Alam}, {Bell}, {Blecic}, {Caceres},
  {Carter}, {Casewell}, {Crouzet}, {Cubillos}, {Decin}, {Fortney}, {Gibson},
  {Heng}, {Henning}, {Iro}, {Kendrew}, {Lagage}, {Leconte}, {Lendl},
  {Lothringer}, {Mancini}, {Mikal-Evans}, {Molaverdikhani}, {Nikolov}, {Ohno},
  {Palle}, {Piaulet}, {Redfield}, {Roy}, {Tsai}, {Venot}, \&
  {Wheatley}}]{coulombe_2023}
{Coulombe}, L.-P., {Benneke}, B., {Challener}, R., {et~al.} 2023, Nature, 620,
  292, \dodoi{10.1038/s41586-023-06230-1}

\bibitem[{{Deming} {et~al.}(1988){Deming}, {Boyle}, {Jennings}, \&
  {Wiedemann}}]{deming_1988}
{Deming}, D., {Boyle}, R.~J., {Jennings}, D.~E., \& {Wiedemann}, G. 1988,
  Ap.J., 333, 978, \dodoi{10.1086/166806}

\bibitem[{{Deming} {et~al.}(2009){Deming}, {Seager}, {Winn}, {Miller-Ricci},
  {Clampin}, {Lindler}, {Greene}, {Charbonneau}, {Laughlin}, {Ricker},
  {Latham}, \& {Ennico}}]{deming_2009}
{Deming}, D., {Seager}, S., {Winn}, J., {et~al.} 2009, PASP, 121, 952,
  \dodoi{10.1086/605913}

\bibitem[{{Deming} {et~al.}(2013){Deming}, {Wilkins}, {McCullough}, {Burrows},
  {Fortney}, {Agol}, {Dobbs-Dixon}, {Madhusudhan}, {Crouzet}, {Desert},
  {Gilliland}, {Haynes}, {Knutson}, {Line}, {Magic}, {Mandell}, {Ranjan},
  {Charbonneau}, {Clampin}, {Seager}, \& {Showman}}]{deming_2013}
{Deming}, D., {Wilkins}, A., {McCullough}, P., {et~al.} 2013, Ap.J., 774, 95,
  \dodoi{10.1088/0004-637X/774/2/95}

\bibitem[{{Dicken} {et~al.}(2022){Dicken}, {Rieke}, {Ressler}, {Morrison},
  {Garcia Marin}, {Argyriou}, {Gordon}, {Regan}, {Cossou}, {Gaspar}, {Glasse},
  {Guillard}, {Labiano}, \& {Wright}}]{dicken_2022}
{Dicken}, D., {Rieke}, G., {Ressler}, M., {et~al.} 2022, in Society of
  Photo-Optical Instrumentation Engineers (SPIE) Conference Series, Vol. 12180,
  Space Telescopes and Instrumentation 2022: Optical, Infrared, and Millimeter
  Wave, ed. L.~E. {Coyle}, S.~{Matsuura}, \& M.~D. {Perrin}, 121802R,
  \dodoi{10.1117/12.2630027}

\bibitem[{{Dicken} {et~al.}(2024){Dicken}, {Garc{\'\i}a Mar{\'\i}n}, {Shivaei},
  {Guillard}, {Libralato}, {Glasse}, {Gordon}, {Cossou}, {Kavanagh}, {Temim},
  {Flagey}, {Klaassen}, {Rieke}, {Wright}, {Alberts}, {Azzollini},
  {{\'A}lvarez-M{\'a}rquez}, {Bouchet}, {Bright}, {Cracraft}, {Coulais}, {Hunor
  Detre}, {Engesser}, {Fox}, {Gaspar}, {Gastaud}, {Glauser}, {Hines},
  {Kendrew}, {Labiano}, {Lagage}, {Lee}, {Law}, {Morrison}, {Noriega-Crespo},
  {Jones}, {Patapis}, {Scheithauer}, {Sloan}, \& {Tamaz}}]{dicken_2024}
{Dicken}, D., {Garc{\'\i}a Mar{\'\i}n}, M., {Shivaei}, I., {et~al.} 2024, arXiv
  e-prints, arXiv:2403.16686, \dodoi{10.48550/arXiv.2403.16686}

\bibitem[{{Doyon} {et~al.}(2012){Doyon}, {Hutchings}, {Beaulieu}, {Albert},
  {Lafreni{\`e}re}, {Willott}, {Touahri}, {Rowlands}, {Maszkiewicz},
  {Fullerton}, {Volk}, {Martel}, {Chayer}, {Sivaramakrishnan}, {Abraham},
  {Ferrarese}, {Jayawardhana}, {Johnstone}, {Meyer}, {Pipher}, \&
  {Sawicki}}]{doyon_2012}
{Doyon}, R., {Hutchings}, J.~B., {Beaulieu}, M., {et~al.} 2012, in Society of
  Photo-Optical Instrumentation Engineers (SPIE) Conference Series, Vol. 8442,
  Space Telescopes and Instrumentation 2012: Optical, Infrared, and Millimeter
  Wave, ed. M.~C. {Clampin}, G.~G. {Fazio}, H.~A. {MacEwen}, \& J.~{Oschmann},
  Jacobus~M., 84422R, \dodoi{10.1117/12.926578}

\bibitem[{{Dyrek} {et~al.}(2024{\natexlab{a}}){Dyrek}, {Ducrot}, {Lagage},
  {Tremblin}, {Kendrew}, {Bouwman}, \& {Bouffet}}]{dyrek_2024a}
{Dyrek}, A., {Ducrot}, E., {Lagage}, P.~O., {et~al.} 2024{\natexlab{a}},
  Astr.Ap., 683, A212, \dodoi{10.1051/0004-6361/202347127}

\bibitem[{{Dyrek} {et~al.}(2024{\natexlab{b}}){Dyrek}, {Min}, {Decin},
  {Bouwman}, {Crouzet}, {Molli{\`e}re}, {Lagage}, {Konings}, {Tremblin},
  {G{\"u}del}, {Pye}, {Waters}, {Henning}, {Vandenbussche}, {Ardevol Martinez},
  {Argyriou}, {Ducrot}, {Heinke}, {van Looveren}, {Absil}, {Barrado}, {Baudoz},
  {Boccaletti}, {Cossou}, {Coulais}, {Edwards}, {Gastaud}, {Glasse}, {Glauser},
  {Greene}, {Kendrew}, {Krause}, {Lahuis}, {Mueller}, {Olofsson}, {Patapis},
  {Rouan}, {Royer}, {Scheithauer}, {Waldmann}, {Whiteford}, {Colina}, {van
  Dishoeck}, {{\"O}stlin}, {Ray}, \& {Wright}}]{dyrek_2024b}
{Dyrek}, A., {Min}, M., {Decin}, L., {et~al.} 2024{\natexlab{b}}, Nature, 625,
  51, \dodoi{10.1038/s41586-023-06849-0}

\bibitem[{{Esparza-Borges} {et~al.}(2023){Esparza-Borges}, {L{\'o}pez-Morales},
  {Adams Redai}, {Pall{\'e}}, {Kirk}, {Casasayas-Barris}, {Batalha}, {Rackham},
  {Bean}, {Casewell}, {Decin}, {Dos Santos}, {Garc{\'\i}a Mu{\~n}oz},
  {Harrington}, {Heng}, {Hu}, {Mancini}, {Molaverdikhani}, {Morello},
  {Nikolov}, {Nixon}, {Redfield}, {Stevenson}, {Wakeford}, {Alam}, {Benneke},
  {Blecic}, {Crouzet}, {Daylan}, {Inglis}, {Kreidberg}, {Petit dit de la
  Roche}, \& {Turner}}]{esparza-borges_2023}
{Esparza-Borges}, E., {L{\'o}pez-Morales}, M., {Adams Redai}, J.~I., {et~al.}
  2023, Ap.J.(Lett.), 955, L19, \dodoi{10.3847/2041-8213/acf27b}

\bibitem[{{Espinoza} {et~al.}(2023){Espinoza}, {{\'U}beda}, {Birkmann},
  {Ferruit}, {Valenti}, {Sing}, {Rustamkulov}, {Regan}, {Kendrew}, {Sabbi},
  {Schlawin}, {Beatty}, {Albert}, {Greene}, {Nikolov}, {Karakla}, {Keyes},
  {Alves de Oliveira}, {B{\"o}ker}, {Pena-Guerrero}, {Giardino}, {Kumari},
  {Manjavacas}, {Proffitt}, \& {Rawle}}]{espinoza_2023}
{Espinoza}, N., {{\'U}beda}, L., {Birkmann}, S.~M., {et~al.} 2023, PASP, 135,
  018002, \dodoi{10.1088/1538-3873/aca3d3}

\bibitem[{{Feinstein} {et~al.}(2023){Feinstein}, {Radica}, {Welbanks},
  {Murray}, {Ohno}, {Coulombe}, {Espinoza}, {Bean}, {Teske}, {Benneke}, {Line},
  {Rustamkulov}, {Saba}, {Tsiaras}, {Barstow}, {Fortney}, {Gao}, {Knutson},
  {MacDonald}, {Mikal-Evans}, {Rackham}, {Taylor}, {Parmentier}, {Batalha},
  {Berta-Thompson}, {Carter}, {Changeat}, {dos Santos}, {Gibson}, {Goyal},
  {Kreidberg}, {L{\'o}pez-Morales}, {Lothringer}, {Miguel}, {Molaverdikhani},
  {Moran}, {Morello}, {Mukherjee}, {Sing}, {Stevenson}, {Wakeford}, {Ahrer},
  {Alam}, {Alderson}, {Allen}, {Batalha}, {Bell}, {Blecic}, {Brande},
  {Caceres}, {Casewell}, {Chubb}, {Crossfield}, {Crouzet}, {Cubillos}, {Decin},
  {D{\'e}sert}, {Harrington}, {Heng}, {Henning}, {Iro}, {Kempton}, {Kendrew},
  {Kirk}, {Krick}, {Lagage}, {Lendl}, {Mancini}, {Mansfield}, {May}, {Mayne},
  {Nikolov}, {Palle}, {Petit dit de la Roche}, {Piaulet}, {Powell}, {Redfield},
  {Rogers}, {Roman}, {Roy}, {Nixon}, {Schlawin}, {Tan}, {Tremblin}, {Turner},
  {Venot}, {Waalkes}, {Wheatley}, \& {Zhang}}]{feinstein_2023}
{Feinstein}, A.~D., {Radica}, M., {Welbanks}, L., {et~al.} 2023, Nature, 614,
  670, \dodoi{10.1038/s41586-022-05674-1}

\bibitem[{{Fu} {et~al.}(2022){Fu}, {Espinoza}, {Sing}, {Lothringer}, {Dos
  Santos}, {Rustamkulov}, {Deming}, {Kempton}, {Komacek}, {Knutson}, {Albert},
  {Pontoppidan}, {Volk}, \& {Filippazzo}}]{Fu_2022}
{Fu}, G., {Espinoza}, N., {Sing}, D.~K., {et~al.} 2022, Ap.J.(Lett.), 940, L35,
  \dodoi{10.3847/2041-8213/ac9977}

\bibitem[{{Fu} {et~al.}(2024){Fu}, {Welbanks}, {Deming}, {Inglis}, {Zhang},
  {Lothringer}, {Ih}, {Moses}, {Schlawin}, {Knutson}, {Henry}, {Greene},
  {Sing}, {Savel}, {Kempton}, {Louie}, {Line}, \& {Nixon}}]{Fu_2024}
{Fu}, G., {Welbanks}, L., {Deming}, D., {et~al.} 2024, arXiv e-prints,
  arXiv:2407.06163, \dodoi{10.48550/arXiv.2407.06163}

\bibitem[{{Gasman} {et~al.}(2023){Gasman}, {Argyriou}, {Sloan}, {Aringer},
  {{\'A}lvarez-M{\'a}rquez}, {Fox}, {Glasse}, {Glauser}, {Jones}, {Justtanont},
  {Kavanagh}, {Klaassen}, {Labiano}, {Larson}, {Law}, {Mueller}, {Nayak},
  {Noriega-Crespo}, {Patapis}, {Royer}, \& {Vandenbussche}}]{gasman_2023}
{Gasman}, D., {Argyriou}, I., {Sloan}, G.~C., {et~al.} 2023, Astr.Ap., 673,
  A102, \dodoi{10.1051/0004-6361/202245633}

\bibitem[{{Gilbert} {et~al.}(2021){Gilbert}, {Barclay}, {Kruse}, {Quintana}, \&
  {Walkowicz}}]{gilbert_2021}
{Gilbert}, E.~A., {Barclay}, T., {Kruse}, E., {Quintana}, E.~V., \&
  {Walkowicz}, L.~M. 2021, Frontiers in Astronomy and Space Sciences, 8, 190,
  \dodoi{10.3389/fspas.2021.769371}

\bibitem[{{Grant} {et~al.}(2023){Grant}, {Lothringer}, {Wakeford}, {Alam},
  {Alderson}, {Bean}, {Benneke}, {D{\'e}sert}, {Daylan}, {Flagg}, {Hu},
  {Inglis}, {Kirk}, {Kreidberg}, {L{\'o}pez-Morales}, {Mancini}, {Mikal-Evans},
  {Molaverdikhani}, {Palle}, {Rackham}, {Redfield}, {Stevenson}, {Valenti},
  {Wallack}, {Aggarwal}, {Ahrer}, {Crossfield}, {Crouzet}, {Iro}, {Nikolov},
  {Wheatley}, \& {JWST Transiting Exoplanet Community ERS Team}}]{grant_2023}
{Grant}, D., {Lothringer}, J.~D., {Wakeford}, H.~R., {et~al.} 2023,
  Ap.J.(Lett.), 949, L15, \dodoi{10.3847/2041-8213/acd544}

\bibitem[{{Greene} {et~al.}(2023){Greene}, {Bell}, {Ducrot}, {Dyrek}, {Lagage},
  \& {Fortney}}]{greene_2023}
{Greene}, T.~P., {Bell}, T.~J., {Ducrot}, E., {et~al.} 2023, \nat, 618, 39,
  \dodoi{10.1038/s41586-023-05951-7}

\bibitem[{{Greene} {et~al.}(2016){Greene}, {Line}, {Montero}, {Fortney},
  {Lustig-Yaeger}, \& {Luther}}]{greene_2016}
{Greene}, T.~P., {Line}, M.~R., {Montero}, C., {et~al.} 2016, Ap.J., 817, 17,
  \dodoi{10.3847/0004-637X/817/1/17}

\bibitem[{{Greene} {et~al.}(2017){Greene}, {Kelly}, {Stansberry}, {Leisenring},
  {Egami}, {Schlawin}, {Chu}, {Hodapp}, \& {Rieke}}]{greene_2017}
{Greene}, T.~P., {Kelly}, D.~M., {Stansberry}, J., {et~al.} 2017, Journal of
  Astronomical Telescopes, Instruments, and Systems, 3, 035001,
  \dodoi{10.1117/1.JATIS.3.3.035001}

\bibitem[{{Guinan}(1977)}]{guinan_1977}
{Guinan}, E.~F. 1977, Astron.J., 82, 51, \dodoi{10.1086/112008}

\bibitem[{{Hauschildt} {et~al.}(1999){Hauschildt}, {Allard}, \&
  {Baron}}]{hauschildt_1999}
{Hauschildt}, P.~H., {Allard}, F., \& {Baron}, E. 1999, Ap.J., 512, 377,
  \dodoi{10.1086/306745}

\bibitem[{{Henning} {et~al.}(2024){Henning}, {Kamp}, {Samland}, {Arabhavi},
  {Kanwar}, {van Dishoeck}, {G{\"u}del}, {Lagage}, {Waelkens}, {Abergel},
  {Absil}, {Barrado}, {Boccaletti}, {Bouwman}, {Caratti o Garatti}, {Geers},
  {Glauser}, {Lahuis}, {Mueller}, {Nehm{\'e}}, {Olofsson}, {Pantin}, {Ray},
  {Scheithauer}, {Vandenbussche}, {Waters}, {Wright}, {Argyriou},
  {Christiaens}, {Franceschi}, {Gasman}, {Grant}, {Guadarrama}, {Jang},
  {Morales-Calder{\'o}n}, {Pawellek}, {Perotti}, {Rodgers-Lee}, {Schreiber},
  {Schwarz}, {Tabone}, {Temmink}, {Vlasblom}, {Colina}, {Greve}, \&
  {{\"O}stlin}}]{henning_2024}
{Henning}, T., {Kamp}, I., {Samland}, M., {et~al.} 2024, PASP, 136, 054302,
  \dodoi{10.1088/1538-3873/ad3455}

\bibitem[{{Holmberg} \& {Madhusudhan}(2023)}]{holmberg_2023}
{Holmberg}, M., \& {Madhusudhan}, N. 2023, MNRAS, 524, 377,
  \dodoi{10.1093/mnras/stad1580}

\bibitem[{{Horne}(1986)}]{horne_1986}
{Horne}, K. 1986, PASP, 98, 609, \dodoi{10.1086/131801}

\bibitem[{{Jakobsen} {et~al.}(2022){Jakobsen}, {Ferruit}, {Alves de Oliveira},
  {Arribas}, {Bagnasco}, {Barho}, {Beck}, {Birkmann}, {B{\"o}ker}, {Bunker},
  {Charlot}, {de Jong}, {de Marchi}, {Ehrenwinkler}, {Falcolini}, {Fels},
  {Franx}, {Franz}, {Funke}, {Giardino}, {Gnata}, {Holota}, {Honnen}, {Jensen},
  {Jentsch}, {Johnson}, {Jollet}, {Karl}, {Kling}, {K{\"o}hler}, {Kolm},
  {Kumari}, {Lander}, {Lemke}, {L{\'o}pez-Caniego}, {L{\"u}tzgendorf},
  {Maiolino}, {Manjavacas}, {Marston}, {Maschmann}, {Maurer}, {Messerschmidt},
  {Moseley}, {Mosner}, {Mott}, {Muzerolle}, {Pirzkal}, {Pittet}, {Plitzke},
  {Posselt}, {Rapp}, {Rauscher}, {Rawle}, {Rix}, {R{\"o}del}, {Rumler},
  {Sabbi}, {Salvignol}, {Schmid}, {Sirianni}, {Smith}, {Strada}, {te Plate},
  {Valenti}, {Wettemann}, {Wiehe}, {Wiesmayer}, {Willott}, {Wright}, {Zeidler},
  \& {Zincke}}]{jakobsen_2022}
{Jakobsen}, P., {Ferruit}, P., {Alves de Oliveira}, C., {et~al.} 2022,
  Astr.Ap., 661, A80, \dodoi{10.1051/0004-6361/202142663}

\bibitem[{{Jenkins} {et~al.}(2019){Jenkins}, {Harrington}, {Challener},
  {Kurtovic}, {Ramirez}, {Pe{\~n}a}, {McIntyre}, {Himes}, {Rodr{\'\i}guez},
  {Anglada-Escud{\'e}}, {Dreizler}, {Ofir}, {Pe{\~n}a Rojas}, {Ribas}, {Rojo},
  {Kipping}, {Butler}, {Amado}, {Rodr{\'\i}guez-L{\'o}pez}, {Kempton}, {Palle},
  \& {Murgas}}]{jenkins_2019}
{Jenkins}, J.~S., {Harrington}, J., {Challener}, R.~C., {et~al.} 2019, MNRAS,
  487, 268, \dodoi{10.1093/mnras/stz1268}

\bibitem[{{Kempton} {et~al.}(2023){Kempton}, {Zhang}, {Bean}, {Steinrueck},
  {Piette}, {Parmentier}, {Malsky}, {Roman}, {Rauscher}, {Gao}, {Bell}, {Xue},
  {Taylor}, {Savel}, {Arnold}, {Nixon}, {Stevenson}, {Mansfield}, {Kendrew},
  {Zieba}, {Ducrot}, {Dyrek}, {Lagage}, {Stassun}, {Henry}, {Barman}, {Lupu},
  {Malik}, {Kataria}, {Ih}, {Fu}, {Welbanks}, \& {McGill}}]{kempton_2023}
{Kempton}, E. M.~R., {Zhang}, M., {Bean}, J.~L., {et~al.} 2023, Nature, 620,
  67, \dodoi{10.1038/s41586-023-06159-5}

\bibitem[{{Kendrew} {et~al.}(2015){Kendrew}, {Scheithauer}, {Bouchet},
  {Amiaux}, {Azzollini}, {Bouwman}, {Chen}, {Dubreuil}, {Fischer}, {Glasse},
  {Greene}, {Lagage}, {Lahuis}, {Ronayette}, {Wright}, \&
  {Wright}}]{kendrew_2015}
{Kendrew}, S., {Scheithauer}, S., {Bouchet}, P., {et~al.} 2015, PASP, 127, 623,
  \dodoi{10.1086/682255}

\bibitem[{{Kester} {et~al.}(2003){Kester}, {Beintema}, \& {Lutz}}]{kester_2003}
{Kester}, D.~J.~M., {Beintema}, D.~A., \& {Lutz}, D. 2003, in ESA Special
  Publication, Vol. 481, The Calibration Legacy of the ISO Mission, ed.
  L.~{Metcalfe}, A.~{Salama}, S.~B. {Peschke}, \& M.~F. {Kessler}, 375

\bibitem[{{Knutson} {et~al.}(2014){Knutson}, {Dragomir}, {Kreidberg},
  {Kempton}, {McCullough}, {Fortney}, {Bean}, {Gillon}, {Homeier}, \&
  {Howard}}]{knutson_2014}
{Knutson}, H.~A., {Dragomir}, D., {Kreidberg}, L., {et~al.} 2014, Ap.J., 794,
  155, \dodoi{10.1088/0004-637X/794/2/155}

\bibitem[{{Kreidberg} \& {Loeb}(2016)}]{kreidberg_2016}
{Kreidberg}, L., \& {Loeb}, A. 2016, Ap.J.(Lett.), 832, L12,
  \dodoi{10.3847/2041-8205/832/1/L12}

\bibitem[{{Labiano} {et~al.}(2016){Labiano}, {Azzollini}, {Bailey}, {Beard},
  {Dicken}, {Garc{\'\i}a-Mar{\'\i}n}, {Geers}, {Glasse}, {Glauser}, {Gordon},
  {Justtanont}, {Klaassen}, {Lahuis}, {Law}, {Morrison}, {M{\"u}ller}, {Rieke},
  {Vandenbussche}, \& {Wright}}]{labiano_2016}
{Labiano}, A., {Azzollini}, R., {Bailey}, J., {et~al.} 2016, in Society of
  Photo-Optical Instrumentation Engineers (SPIE) Conference Series, Vol. 9910,
  Observatory Operations: Strategies, Processes, and Systems VI, ed. A.~B.
  {Peck}, R.~L. {Seaman}, \& C.~R. {Benn}, 99102W, \dodoi{10.1117/12.2232554}

\bibitem[{{Labiano} {et~al.}(2021){Labiano}, {Argyriou},
  {{\'A}lvarez-M{\'a}rquez}, {Glasse}, {Glauser}, {Patapis}, {Law}, {Brandl},
  {Justtanont}, {Lahuis}, {Mart{\'\i}nez-Galarza}, {Mueller}, {Noriega-Crespo},
  {Royer}, {Shaughnessy}, \& {Vandenbussche}}]{labiano_2021}
{Labiano}, A., {Argyriou}, I., {{\'A}lvarez-M{\'a}rquez}, J., {et~al.} 2021,
  Astr.Ap., 656, A57, \dodoi{10.1051/0004-6361/202140614}

\bibitem[{{Law} {et~al.}(2023){Law}, {E. Morrison}, {Argyriou}, {Patapis},
  {{\'A}lvarez-M{\'a}rquez}, {Labiano}, \& {Vandenbussche}}]{law_2023}
{Law}, D.~R., {E. Morrison}, J., {Argyriou}, I., {et~al.} 2023, Astron.J., 166,
  45, \dodoi{10.3847/1538-3881/acdddc}

\bibitem[{{Lehmann} {et~al.}(2018){Lehmann}, {Tsymbal}, {Pertermann},
  {Tkachenko}, {Mkrtichian}, \& {A-thano}}]{lehmann_2018}
{Lehmann}, H., {Tsymbal}, V., {Pertermann}, F., {et~al.} 2018, \aap, 615, A131,
  \dodoi{10.1051/0004-6361/201629914}

\bibitem[{{Limbach} {et~al.}(2022){Limbach}, {Vanderburg}, {Stevenson},
  {Blouin}, {Morley}, {Lustig-Yaeger}, {Soares-Furtado}, \&
  {Janson}}]{limbach_2022}
{Limbach}, M.~A., {Vanderburg}, A., {Stevenson}, K.~B., {et~al.} 2022, MNRAS,
  517, 2622, \dodoi{10.1093/mnras/stac2823}

\bibitem[{{Lustig-Yaeger} {et~al.}(2023){Lustig-Yaeger}, {Fu}, {May},
  {Ceballos}, {Moran}, {Peacock}, {Stevenson}, {Kirk}, {L{\'o}pez-Morales},
  {MacDonald}, {Mayorga}, {Sing}, {Sotzen}, {Valenti}, {Redai}, {Alam},
  {Batalha}, {Bennett}, {Gonzalez-Quiles}, {Kruse}, {Lothringer},
  {Rustamkulov}, \& {Wakeford}}]{lustig-yaeger_2023}
{Lustig-Yaeger}, J., {Fu}, G., {May}, E.~M., {et~al.} 2023, Nature Astronomy,
  7, 1317, \dodoi{10.1038/s41550-023-02064-z}

\bibitem[{{Madhusudhan} {et~al.}(2023){Madhusudhan}, {Sarkar}, {Constantinou},
  {Holmberg}, {Piette}, \& {Moses}}]{madhusudhan_2023}
{Madhusudhan}, N., {Sarkar}, S., {Constantinou}, S., {et~al.} 2023,
  Ap.J.(Lett.), 956, L13, \dodoi{10.3847/2041-8213/acf577}

\bibitem[{{Mikal-Evans} {et~al.}(2023){Mikal-Evans}, {Sing}, {Dong},
  {Foreman-Mackey}, {Kataria}, {Barstow}, {Goyal}, {Lewis}, {Lothringer},
  {Mayne}, {Wakeford}, {Christie}, \& {Rustamkulov}}]{mikal-evans_2023}
{Mikal-Evans}, T., {Sing}, D.~K., {Dong}, J., {et~al.} 2023, Ap.J.(Lett.), 943,
  L17, \dodoi{10.3847/2041-8213/acb049}

\bibitem[{{Miles} {et~al.}(2023){Miles}, {Biller}, {Patapis}, {Worthen},
  {Rickman}, {Hoch}, {Skemer}, {Perrin}, {Whiteford}, {Chen}, {Sargent},
  {Mukherjee}, {Morley}, {Moran}, {Bonnefoy}, {Petrus}, {Carter}, {Choquet},
  {Hinkley}, {Ward-Duong}, {Leisenring}, {Millar-Blanchaer}, {Pueyo}, {Ray},
  {Sallum}, {Stapelfeldt}, {Stone}, {Wang}, {Absil}, {Balmer}, {Boccaletti},
  {Bonavita}, {Booth}, {Bowler}, {Chauvin}, {Christiaens}, {Currie},
  {Danielski}, {Fortney}, {Girard}, {Grady}, {Greenbaum}, {Henning}, {Hines},
  {Janson}, {Kalas}, {Kammerer}, {Kennedy}, {Kenworthy}, {Kervella}, {Lagage},
  {Lew}, {Liu}, {Macintosh}, {Marino}, {Marley}, {Marois}, {Matthews},
  {Matthews}, {Mawet}, {McElwain}, {Metchev}, {Meyer}, {Molliere}, {Pantin},
  {Quirrenbach}, {Rebollido}, {Ren}, {Schneider}, {Vasist}, {Wyatt}, {Zhou},
  {Briesemeister}, {Bryan}, {Calissendorff}, {Cantalloube}, {Cugno}, {De
  Furio}, {Dupuy}, {Factor}, {Faherty}, {Fitzgerald}, {Franson}, {Gonzales},
  {Hood}, {Howe}, {Kraus}, {Kuzuhara}, {Lagrange}, {Lawson}, {Lazzoni}, {Liu},
  {Llop-Sayson}, {Lloyd}, {Martinez}, {Mazoyer}, {Quanz}, {Redai}, {Samland},
  {Schlieder}, {Tamura}, {Tan}, {Uyama}, {Vigan}, {Vos}, {Wagner}, {Wolff},
  {Ygouf}, {Zhang}, {Zhang}, \& {Zhang}}]{miles_2023}
{Miles}, B.~E., {Biller}, B.~A., {Patapis}, P., {et~al.} 2023, Ap.J.(Lett.),
  946, L6, \dodoi{10.3847/2041-8213/acb04a}

\bibitem[{{Moran} {et~al.}(2023){Moran}, {Stevenson}, {Sing}, {MacDonald},
  {Kirk}, {Lustig-Yaeger}, {Peacock}, {Mayorga}, {Bennett},
  {L{\'o}pez-Morales}, {May}, {Rustamkulov}, {Valenti}, {Adams Redai}, {Alam},
  {Batalha}, {Fu}, {Gonzalez-Quiles}, {Highland}, {Kruse}, {Lothringer}, {Ortiz
  Ceballos}, {Sotzen}, \& {Wakeford}}]{moran_2023}
{Moran}, S.~E., {Stevenson}, K.~B., {Sing}, D.~K., {et~al.} 2023, Ap.J.(Lett.),
  948, L11, \dodoi{10.3847/2041-8213/accb9c}

\bibitem[{{Morrison} {et~al.}(2023){Morrison}, {Dicken}, {Argyriou}, {Ressler},
  {Gordon}, {Regan}, {Cracraft}, {Rieke}, {Engesser}, {Alberts},
  {Alvarez-Marquez}, {Colbert}, {Fox}, {Gasman}, {Law}, {Garcia Marin},
  {G{\'a}sp{\'a}r}, {Guillard}, {Kendrew}, {Labiano}, {Laine},
  {Noriega-Crespo}, {Shivaei}, \& {Sloan}}]{morrison_2023}
{Morrison}, J.~E., {Dicken}, D., {Argyriou}, I., {et~al.} 2023, \pasp, 135,
  075004, \dodoi{10.1088/1538-3873/acdea6}

\bibitem[{{Patapis} {et~al.}(2022){Patapis}, {Nasedkin}, {Cugno}, {Glauser},
  {Argyriou}, {Whiteford}, {Molli{\`e}re}, {Glasse}, \& {Quanz}}]{patapis_2022}
{Patapis}, P., {Nasedkin}, E., {Cugno}, G., {et~al.} 2022, Astr.Ap., 658, A72,
  \dodoi{10.1051/0004-6361/202141663}

\bibitem[{{Radica} {et~al.}(2023){Radica}, {Welbanks}, {Espinoza}, {Taylor},
  {Coulombe}, {Feinstein}, {Goyal}, {Scarsdale}, {Albert}, {Baghel}, {Bean},
  {Blecic}, {Lafreni{\`e}re}, {MacDonald}, {Zamyatina}, {Allart1}, {Artigau},
  {Batalha}, {Cook}, {Cowan}, {Dang}, {Doyon}, {Fournier-Tondreau},
  {Johnstone}, {Line}, {Moran}, {Mukherjee}, {Pelletier}, {Roy}, {Talens},
  {Filippazzo}, {Pontoppidan}, \& {Volk}}]{radica_2023}
{Radica}, M., {Welbanks}, L., {Espinoza}, N., {et~al.} 2023, MNRAS, 524, 835,
  \dodoi{10.1093/mnras/stad1762}

\bibitem[{{Ressler} {et~al.}(2015){Ressler}, {Sukhatme}, {Franklin}, {Mahoney},
  {Thelen}, {Bouchet}, {Colbert}, {Cracraft}, {Dicken}, {Gastaud}, {Goodson},
  {Eccleston}, {Moreau}, {Rieke}, \& {Schneider}}]{ressler_2015}
{Ressler}, M.~E., {Sukhatme}, K.~G., {Franklin}, B.~R., {et~al.} 2015, PASP,
  127, 675, \dodoi{10.1086/682258}

\bibitem[{{Ricker} {et~al.}(2015){Ricker}, {Winn}, {Vanderspek}, {Latham},
  {Bakos}, {Bean}, {Berta-Thompson}, {Brown}, {Buchhave}, {Butler}, {Butler},
  {Chaplin}, {Charbonneau}, {Christensen-Dalsgaard}, {Clampin}, {Deming},
  {Doty}, {De Lee}, {Dressing}, {Dunham}, {Endl}, {Fressin}, {Ge}, {Henning},
  {Holman}, {Howard}, {Ida}, {Jenkins}, {Jernigan}, {Johnson}, {Kaltenegger},
  {Kawai}, {Kjeldsen}, {Laughlin}, {Levine}, {Lin}, {Lissauer}, {MacQueen},
  {Marcy}, {McCullough}, {Morton}, {Narita}, {Paegert}, {Palle}, {Pepe},
  {Pepper}, {Quirrenbach}, {Rinehart}, {Sasselov}, {Sato}, {Seager},
  {Sozzetti}, {Stassun}, {Sullivan}, {Szentgyorgyi}, {Torres}, {Udry}, \&
  {Villasenor}}]{ricker_2015}
{Ricker}, G.~R., {Winn}, J.~N., {Vanderspek}, R., {et~al.} 2015, Journal of
  Astronomical Telescopes, Instruments, and Systems, 1, 014003,
  \dodoi{10.1117/1.JATIS.1.1.014003}

\bibitem[{{Rigby} {et~al.}(2023){Rigby}, {Lightsey}, {Garc{\'\i}a Mar{\'\i}n},
  {Bowers}, {Smith}, {Glasse}, {McElwain}, {Rieke}, {Chary}, {Liu}, {Clampin},
  {Kimble}, {Kinzel}, {Laidler}, {Mehalick}, {Noriega-Crespo}, {Shivaei},
  {Skelton}, {Stark}, {Temim}, {Wei}, \& {Willott}}]{rigby_2023}
{Rigby}, J.~R., {Lightsey}, P.~A., {Garc{\'\i}a Mar{\'\i}n}, M., {et~al.} 2023,
  PASP, 135, 048002, \dodoi{10.1088/1538-3873/acbcf4}

\bibitem[{{Rustamkulov} {et~al.}(2023){Rustamkulov}, {Sing}, {Mukherjee},
  {May}, {Kirk}, {Schlawin}, {Line}, {Piaulet}, {Carter}, {Batalha}, {Goyal},
  {L{\'o}pez-Morales}, {Lothringer}, {MacDonald}, {Moran}, {Stevenson},
  {Wakeford}, {Espinoza}, {Bean}, {Batalha}, {Benneke}, {Berta-Thompson},
  {Crossfield}, {Gao}, {Kreidberg}, {Powell}, {Cubillos}, {Gibson}, {Leconte},
  {Molaverdikhani}, {Nikolov}, {Parmentier}, {Roy}, {Taylor}, {Turner},
  {Wheatley}, {Aggarwal}, {Ahrer}, {Alam}, {Alderson}, {Allen}, {Banerjee},
  {Barat}, {Barrado}, {Barstow}, {Bell}, {Blecic}, {Brande}, {Casewell},
  {Changeat}, {Chubb}, {Crouzet}, {Daylan}, {Decin}, {D{\'e}sert},
  {Mikal-Evans}, {Feinstein}, {Flagg}, {Fortney}, {Harrington}, {Heng}, {Hong},
  {Hu}, {Iro}, {Kataria}, {Kempton}, {Krick}, {Lendl}, {Lillo-Box}, {Louca},
  {Lustig-Yaeger}, {Mancini}, {Mansfield}, {Mayne}, {Miguel}, {Morello},
  {Ohno}, {Palle}, {Petit dit de la Roche}, {Rackham}, {Radica},
  {Ramos-Rosado}, {Redfield}, {Rogers}, {Shkolnik}, {Southworth}, {Teske},
  {Tremblin}, {Tucker}, {Venot}, {Waalkes}, {Welbanks}, {Zhang}, \&
  {Zieba}}]{rustamkulov_2023}
{Rustamkulov}, Z., {Sing}, D.~K., {Mukherjee}, S., {et~al.} 2023, Nature, 614,
  659, \dodoi{10.1038/s41586-022-05677-y}

\bibitem[{{Sawyer}(1887)}]{sawyer_1887}
{Sawyer}, E.~F. 1887, Astron.J., 7, 119, \dodoi{10.1086/100933}

\bibitem[{{Schlawin} {et~al.}(2023){Schlawin}, {Beatty}, {Brooks}, {Nikolov},
  {Greene}, {Espinoza}, {Glidic}, {Baka}, {Egami}, {Stansberry}, {Boyer},
  {Gennaro}, {Leisenring}, {Hilbert}, {Misselt}, {Kelly}, {Canipe}, {Beichman},
  {Correnti}, {Knight}, {Jurling}, {Perrin}, {Feinberg}, {McElwain}, {Bond},
  {Ciardi}, {Kendrew}, \& {Rieke}}]{schlawin_2023}
{Schlawin}, E., {Beatty}, T., {Brooks}, B., {et~al.} 2023, PASP, 135, 018001,
  \dodoi{10.1088/1538-3873/aca718}

\bibitem[{{Schwarz}(1978)}]{schwarz_1978}
{Schwarz}, G. 1978, Annals of Statistics, 6, 461

\bibitem[{{Sellek} {et~al.}(2024){Sellek}, {Bajaj}, {Pascucci}, {Clarke},
  {Alexander}, {Xie}, {Ballabio}, {Deng}, {Gorti}, {Gaspar}, \&
  {Morrison}}]{sellek_2024}
{Sellek}, A.~D., {Bajaj}, N.~S., {Pascucci}, I., {et~al.} 2024, Astron.J., 167,
  223, \dodoi{10.3847/1538-3881/ad34ae}

\bibitem[{{Snellen} {et~al.}(2010){Snellen}, {de Kok}, {de Mooij}, \&
  {Albrecht}}]{snellen_2010}
{Snellen}, I. A.~G., {de Kok}, R.~J., {de Mooij}, E. J.~W., \& {Albrecht}, S.
  2010, Nature, 465, 1049, \dodoi{10.1038/nature09111}

\bibitem[{{Snellen} {et~al.}(2017){Snellen}, {D{\'e}sert}, {Waters},
  {Robinson}, {Meadows}, {van Dishoeck}, {Brandl}, {Henning}, {Bouwman},
  {Lahuis}, {Min}, {Lovis}, {Dominik}, {Van Eylen}, {Sing},
  {Anglada-Escud{\'e}}, {Birkby}, \& {Brogi}}]{snellen_2017}
{Snellen}, I.~A.~G., {D{\'e}sert}, J.~M., {Waters}, L.~B.~F.~M., {et~al.} 2017,
  Astron.J., 154, 77, \dodoi{10.3847/1538-3881/aa7fbc}

\bibitem[{{Stevenson} {et~al.}(2014){Stevenson}, {Bean}, {Seifahrt},
  {D{\'e}sert}, {Madhusudhan}, {Bergmann}, {Kreidberg}, \&
  {Homeier}}]{stevenson_2014}
{Stevenson}, K.~B., {Bean}, J.~L., {Seifahrt}, A., {et~al.} 2014, Astron.J.,
  147, 161, \dodoi{10.1088/0004-6256/147/6/161}

\bibitem[{{Storey} \& {Hummer}(1995)}]{storey_1995}
{Storey}, P.~J., \& {Hummer}, D.~G. 1995, MNRAS, 272, 41,
  \dodoi{10.1093/mnras/272.1.41}

\bibitem[{{Taylor} {et~al.}(2023){Taylor}, {Radica}, {Welbanks}, {MacDonald},
  {Blecic}, {Zamyatina}, {Roth}, {Bean}, {Parmentier}, {Coulombe}, {Feinstein},
  {Espinoza}, {Benneke}, {Lafreni{\`e}re}, {Doyon}, \& {Ahrer}}]{taylor_2023}
{Taylor}, J., {Radica}, M., {Welbanks}, L., {et~al.} 2023, MNRAS, 524, 817,
  \dodoi{10.1093/mnras/stad1547}

\bibitem[{{Tomkin}(1985)}]{tomkin_1985}
{Tomkin}, J. 1985, Ap.J., 297, 250, \dodoi{10.1086/163522}

\bibitem[{{Tomkin} \& {Lambert}(1989)}]{tomkin_1989}
{Tomkin}, J., \& {Lambert}, D.~L. 1989, MNRAS, 241, 777,
  \dodoi{10.1093/mnras/241.4.777}

\bibitem[{{Varricatt} \& {Ashok}(1999)}]{varricatt_1999}
{Varricatt}, W.~P., \& {Ashok}, N.~M. 1999, Astron.J., 117, 2980,
  \dodoi{10.1086/300870}

\bibitem[{{Vernazza} {et~al.}(1976){Vernazza}, {Avrett}, \&
  {Loeser}}]{vernazza_1976}
{Vernazza}, J.~E., {Avrett}, E.~H., \& {Loeser}, R. 1976, Ap.J.(Suppl.), 30, 1,
  \dodoi{10.1086/190356}

\bibitem[{{Wakeford} \& {Sing}(2015)}]{wakeford_2015}
{Wakeford}, H.~R., \& {Sing}, D.~K. 2015, Astr.Ap., 573, A122,
  \dodoi{10.1051/0004-6361/201424207}

\bibitem[{{Wells} {et~al.}(2015){Wells}, {Pel}, {Glasse}, {Wright},
  {Aitink-Kroes}, {Azzollini}, {Beard}, {Brandl}, {Gallie}, {Geers}, {Glauser},
  {Hastings}, {Henning}, {Jager}, {Justtanont}, {Kruizinga}, {Lahuis}, {Lee},
  {Martinez-Delgado}, {Mart{\'\i}nez-Galarza}, {Meijers}, {Morrison},
  {M{\"u}ller}, {Nakos}, {O'Sullivan}, {Oudenhuysen}, {Parr-Burman}, {Pauwels},
  {Rohloff}, {Schmalzl}, {Sykes}, {Thelen}, {van Dishoeck}, {Vandenbussche},
  {Venema}, {Visser}, {Waters}, \& {Wright}}]{Wells_2015}
{Wells}, M., {Pel}, J.~W., {Glasse}, A., {et~al.} 2015, PASP, 127, 646,
  \dodoi{10.1086/682281}

\bibitem[{{Wiedemann} {et~al.}(2001){Wiedemann}, {Deming}, \&
  {Bjoraker}}]{wiedeman_2001}
{Wiedemann}, G., {Deming}, D., \& {Bjoraker}, G. 2001, Ap.J., 546, 1068,
  \dodoi{10.1086/318316}

\bibitem[{{Wilson} \& {Devinney}(1971)}]{wilson_1971}
{Wilson}, R.~E., \& {Devinney}, E.~J. 1971, Ap.J., 166, 605,
  \dodoi{10.1086/150986}

\bibitem[{{Wright} {et~al.}(2023){Wright}, {Rieke}, {Glasse}, {Ressler},
  {Garc{\'\i}a Mar{\'\i}n}, {Aguilar}, {Alberts}, {{\'A}lvarez-M{\'a}rquez},
  {Argyriou}, {Banks}, {Baudoz}, {Boccaletti}, {Bouchet}, {Bouwman}, {Brandl},
  {Breda}, {Bright}, {Cale}, {Colina}, {Cossou}, {Coulais}, {Cracraft}, {De
  Meester}, {Dicken}, {Engesser}, {Etxaluze}, {Fox}, {Friedman}, {Fu},
  {Gasman}, {G{\'a}sp{\'a}r}, {Gastaud}, {Geers}, {Glauser}, {Gordon},
  {Greene}, {Greve}, {Grundy}, {G{\"u}del}, {Guillard}, {Haderlein},
  {Hashimoto}, {Henning}, {Hines}, {Holler}, {Detre}, {Jahromi}, {James},
  {Jones}, {Justtanont}, {Kavanagh}, {Kendrew}, {Klaassen}, {Krause},
  {Labiano}, {Lagage}, {Lambros}, {Larson}, {Law}, {Lee}, {Libralato}, {Lorenzo
  Alverez}, {Meixner}, {Morrison}, {Mueller}, {Murray}, {Mycroft}, {Myers},
  {Nayak}, {Naylor}, {Nickson}, {Noriega-Crespo}, {{\"O}stlin}, {O'Sullivan},
  {Ottens}, {Patapis}, {Penanen}, {Pietraszkiewicz}, {Ray}, {Regan},
  {Roteliuk}, {Royer}, {Samara-Ratna}, {Samuelson}, {Sargent}, {Scheithauer},
  {Schneider}, {Schreiber}, {Shaughnessy}, {Sheehan}, {Shivaei}, {Sloan},
  {Tamas}, {Teague}, {Temim}, {Tikkanen}, {Tustain}, {van Dishoeck},
  {Vandenbussche}, {Weilert}, {Whitehouse}, \& {Wolff}}]{wright_2023}
{Wright}, G.~S., {Rieke}, G.~H., {Glasse}, A., {et~al.} 2023, PASP, 135,
  048003, \dodoi{10.1088/1538-3873/acbe66}

\end{thebibliography}
\end{document}